\documentclass[a4paper,11pt]{article}
\usepackage{jinstpub} % for details on the use of the package, please
\usepackage{graphicx}
\usepackage{subfigure}
\usepackage{multirow}
\usepackage{color}
\usepackage{makecell}
\newcommand{\tabincell}[2]{\begin{tabular}{@{}#1@{}}#2\end{tabular}}

\usepackage{lineno}
\title{\boldmath Construction and Simulation Bias Study of The Guide Tube Calibration System for JUNO}

\author[a]{Yuhang Guo,}
\author[a]{Kangfu Zhu,}
\author[a,1]{Qingmin Zhang,\note{Corresponding author.}}
\author[b]{Feiyang Zhang,}
\author[b]{Yue Meng,}
\author[b]{Jianglai Liu,}
\author[a]{Eryuan Qu,}
\affiliation[a]{School of Nuclear Science and Technology, Xi'an Jiaotong University, Xi'an 710049, China}
\affiliation[b]{Shanghai Jiao Tong University, Shanghai 200240, China}
\emailAdd{zhangqingmin@mail.xjtu.edu.cn}

\abstract{
A Guide Tube Calibration System (GTCS) has been designed for the Jiangmen Underground Neutrino Observatory (JUNO), in order to measure the detector energy response near the outer radius of the active volume. Recently, a prototype system has been constructed and tested, and the calibration algorithm has also been studied to evaluate the risk when the simulation tuning and the error control fail. 
In this paper, we first report its construction and the performance tests in the lab. Then the influence on the global energy measurement caused by the simulation bias of GTCS is discussed, in order to make sure the algorithm is qualified.
}


\begin{document}
\maketitle
\flushbottom
\def\degree{${}^{\circ}$}

\section{Introduction}
\label{sec:intro}

The Jiangmen Underground Neutrino Observatory (JUNO) is under construction in South China. It is designed to primarily determine the neutrino mass order (MO) by detecting the inverse beta decay of the anti-neutrinos from both Taishan and Yangjiang Nuclear Power Plants (NPP) at a distance of 53 km. 
Its Central Detector (CD) is an acrylic sphere with an inner diameter of 35.4 m and acrylic thickness of 12 cm, filled with 20 ktons  high light yield (\textasciitilde 1260 photon electrons per MeV) Liquid Scintillator (LS) and equipped with around 18000 20-inch PMTs and 25000 3-inch PMTs. A chimney is constructed on the top of the acrylic sphere for LS filling. Totally 590 stainless steel connection bars are fixed on the acrylic sphere for support. The connection bars are 1.525 m long and 6 cm in diameter and therefore will shadow photons from getting PMTs. All PMTs are immersed in a $\phi 50 m \times 50 m$ water pool. In order to tag the the background due to cosmic muons, a Top Tracker (TT) is designed and deployed above the water pool. \cite{a}

The \textasciitilde 2 to 10 MeV anti-neutrinos from the NPP reactors will react with the protons in the LS and produce correlated pairs of positrons and neutrons. By measuring the light pulse produced by the positron, JUNO is able to measure the energy of anti-neutrinos which will help to determine the MO. Thus, it is crucial to measure the energy accurately. 
With such a superior LS  and high PMT coverage (\textasciitilde 78 \%), JUNO aims to achieve an energy resolution of $3\%/\sqrt{E}$ and energy scale uncertainty of less than 1\%, so that it will be able to determine the neutrino mass order with 90\% C.L. in 6 years data taking. \cite{a,b,c,h,n}

%\begin{figure}[]
%    \centering
%    \includegraphics[width=0.5\textwidth]{figure/Calibration_System.jpg}
%	\caption{Diagram of JUNO Calibration System including the Automatic Calibration Unit (ACU), the Cable Loop System (CLS), the Guide Tube Calibration System (GTCS) and the Remotely-Operated Vehicle (ROV). These four systems will be used for the non-uniformity correction. \textcolor{red}{to be updated and reference to the other subsystems} }
%	\label{fig:Calibration}
%\end{figure}

\begin{figure*}[hb]
    \centering
    \subfigure[Non-uniformity in radial direction.]{    
		\label{fig_CA}     
		\includegraphics[height=0.3\textwidth]{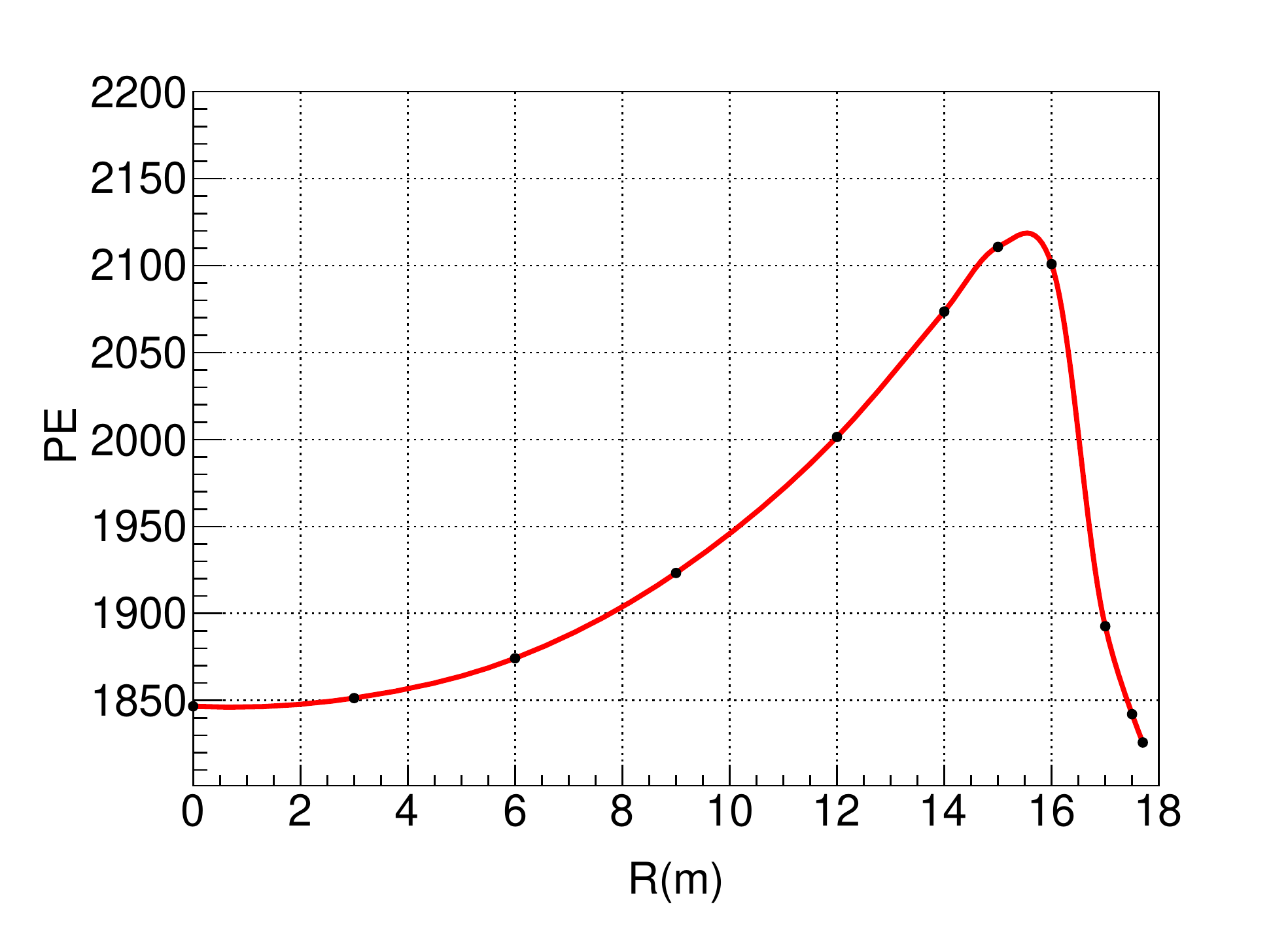}}
    \subfigure[Non-uniformity in polar direction. ]{    
		\label{fig_CB}     
		\includegraphics[height=0.3\textwidth]{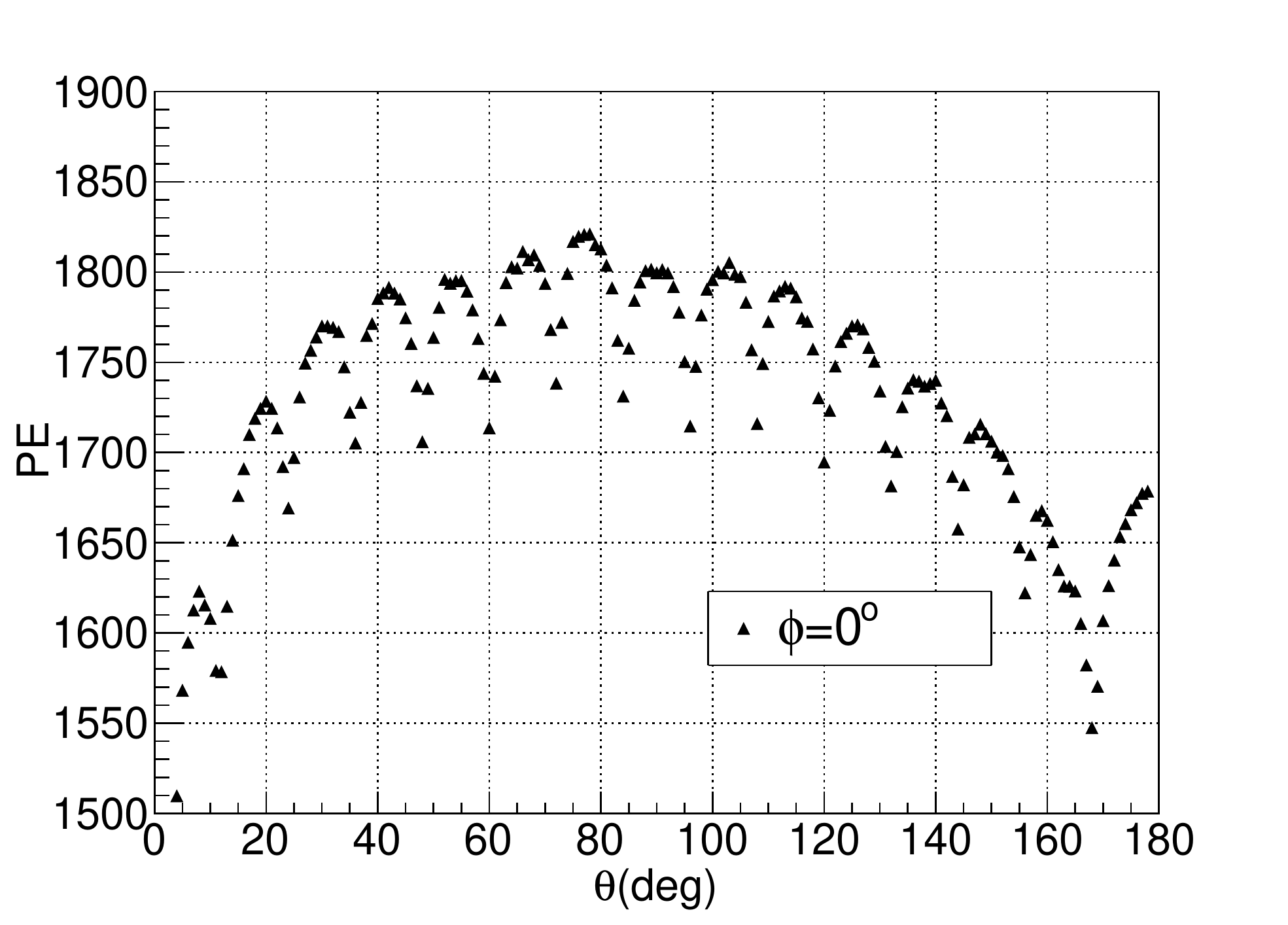}}
    \subfigure[Diagram of JUNO Calibration System. (not to scale)]{       \centering
        \label{fig_CC}     
	    \includegraphics[height=0.5\textwidth]{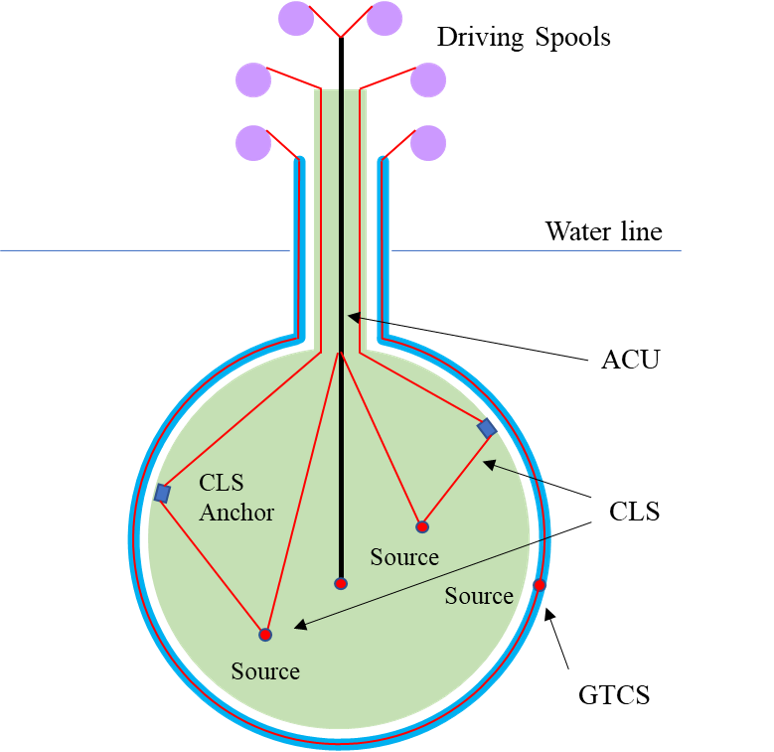}}
	\caption{Energy response non-uniformity of JUNO detector and the diagram of JUNO calibration system. $^{40}K$ is used to plot the energy response non-uniformity. In order to correct it, a complex of calibration systems have been designed, including ACU, CLS and GTCS.}
	\label{fig:Calibration}
\end{figure*}

As shown in Fig.\ref{fig_CA} and Fig.\ref{fig_CB}, due to the geometrical effects and energy leakage, the energy response of JUNO which is the number of photon electrons (PE) recorded by the PMTs, is obviously position dependent inside the detector. 
In the inner space of LS (when R < 15.8 m ), the energy response will rise by around 10 \% as the radius of source position increases since the source is closer to PMT. When the radius of source position is larger than 15.8 m, the energy response will descend due to the total reflection. Structures outside the CD like the chimney or connection bars can shadow some PMTs. Therefore, the energy response varies in both the radial and polar directions. 
In order to correct this energy non-uniformity, a calibration system has been designed \cite{a,Calib_coll}, which is mainly composed of the Auto Calibration Unit (ACU), Cable Loop System (CLS) and Guide Tube Calibration System (GTCS). As shown in Fig.\ref{fig_CC}, The CLS is designed to measure the energy response at designated points in a vertical plane. The ACU and the GTCS measure the energy response along the central vertical axis and at outermost edge of the detector, which is uncovered by the CLS system. A global, continuous 2-D non-uniformity correction function can be constructed by fitting the energy responses at the designated calibration points \cite{Calib_coll}.
In these procedures, the GTCS, which delivers a radioactive source to the desired positions along a longitude, will provide the boundary condition for the fitting. \cite{GT_1st}

In reference \cite{GT_1st}, we have reported the conceptual design, including the calibration strategy and mechanical solution. Recently, the whole system has been further developed and constructed based on the conceptual design. Verification tests of the full system were made in the lab. Considering the huge dimensions and system reliability, it was found to be difficult to make everything run synchronously and safely. Some practical issues, such as the friction, the positioning and the cable threading, have to be addressed and solved before the installation. 
Meanwhile, since the GTCS provides boundary condition for the energy response non-uniformity correction, it is also important to determine how sensitive the global energy measurement is to the simulation bias of GTCS in that correction. This will be helpful to the simulation tuning and error control. 
These issues will be reported in details in this paper.

\section{Construction of GTCS}

\begin{figure}[htp]
    \centering
    \includegraphics[width=0.5\textwidth]{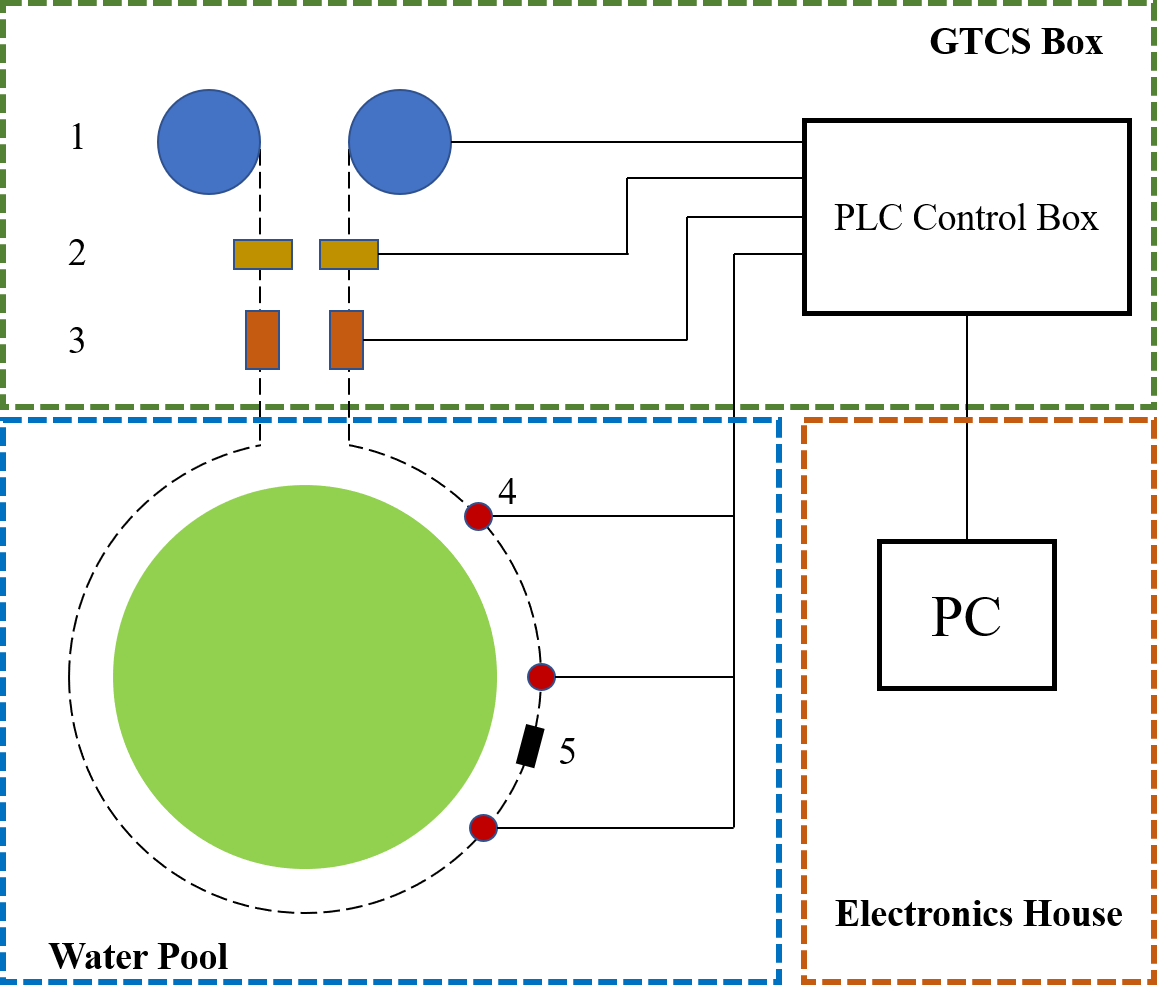}
	\caption{The diagram of GTCS. The guide tube is immersed under water. The source is navigated by the control system, which is in the GTCS box above water. 1.Winding machines; 2. Tension maintainers; 3. Tension sensors; 4. Position sensors. 5. Radioactive source. The whole system can be controlled remotely via a computer in the electronics house. (The components shown are not to scale.)}
	\label{fig:GTCSgeneral}
\end{figure}

A complete system has been designed in order to deliver a radioactive source on CD surface and calibrate the energy response of the JUNO detector. 
The source is pulled back and forth through a guide tube by cables on either end. The tube is fixed on the outer surface of CD to hold the source and the cables. Two winding machines with spools are designed at each end of the tube with cables tied on the spools. One end the cable is unwound from a winding spool as it is wound onto a separate winding spool on the other end. The movement of the spools must be synchronized to minimize cable tension while a positive tension must be maintained to promote consistent winding onto and off the spools.

As shown in Fig.\ref{fig:GTCSgeneral}, the GTCS can be divided into two parts: the guide tube system under water and the control system in the GTCS box. 

\subsection{The guide tube system}

\begin{figure*}[]
    \centering
    \subfigure[]{    
		\label{fig_s}     
	    \includegraphics[height=0.2\textwidth]{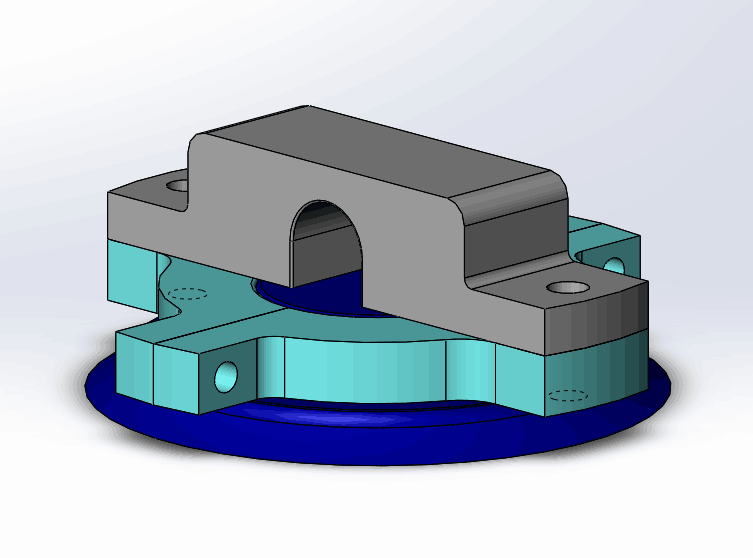}}
    \subfigure[]{    
		\label{fig_t}     
		\includegraphics[height=0.2\textwidth]{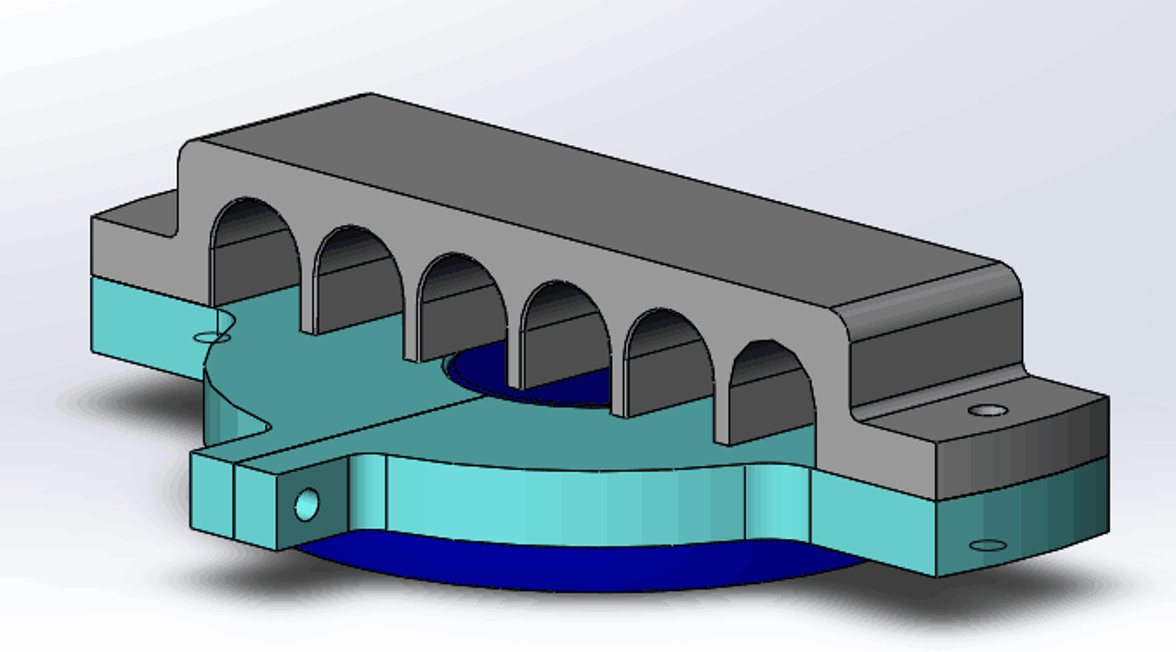}}
    \subfigure[]{    
		\label{fig_u}     
		\includegraphics[height=0.2\textwidth]{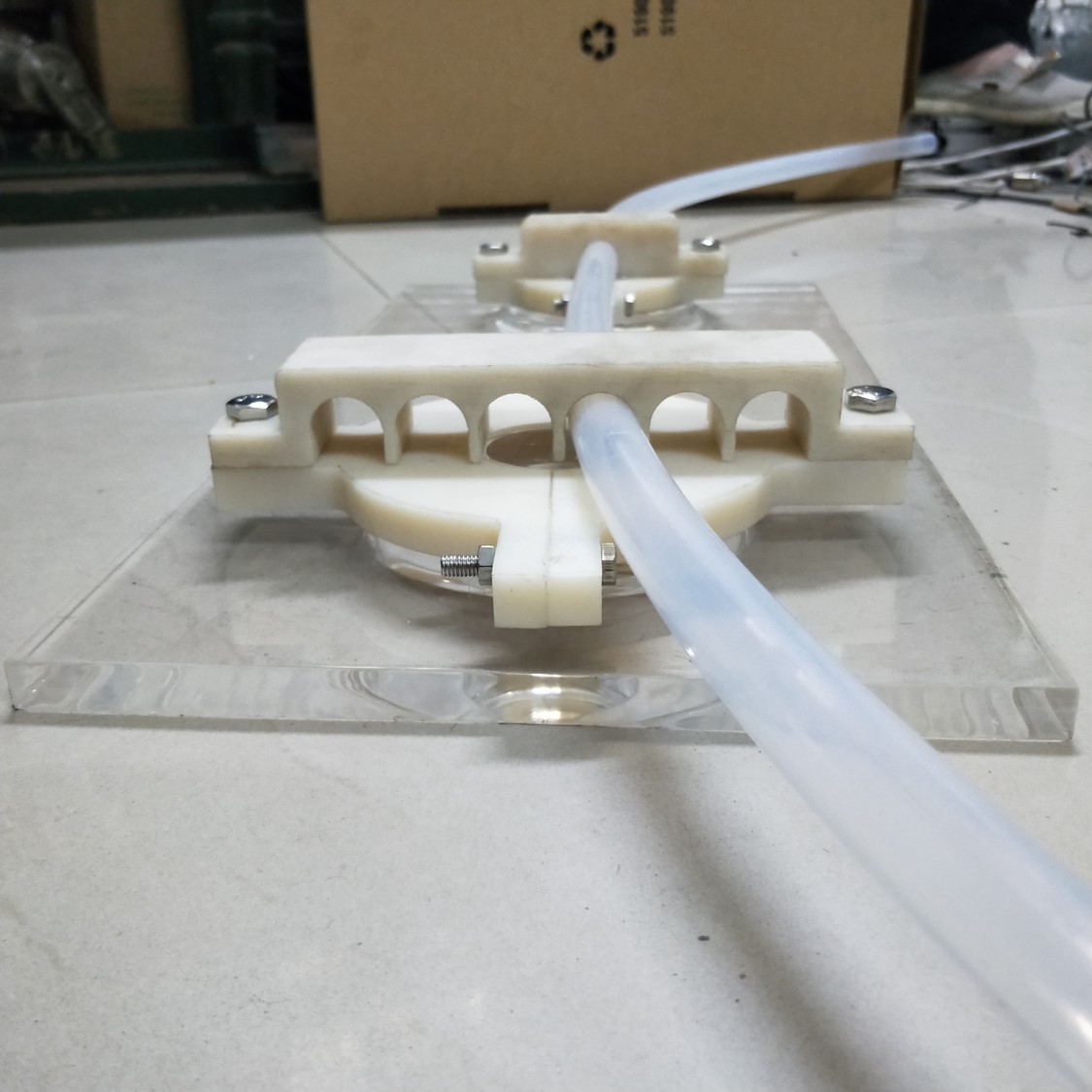}}
	\caption{Two types of GT anchors. (a) is the basic type. (b) is the special designed type in case of GT anchor installation error > 2 cm in horizontal direction. An anchor installation errors smaller than 7 cm can be tuned in this design. (c) Photo with two types of anchors and the PTFE tube. The diameter of the acrylic base (the dark blue one) is 10 cm.}
	\label{fig:GTanchor}
\end{figure*}

The main part of GTCS is a 130 m long PTFE guide tube (GT) with outer-diameter of 19.05 mm and inner-diameter of 16 mm. To fix the tube, 56 anchors will be bonded on the CD outer surface. As demonstrated in Fig.\ref{fig:GTanchor}, two types of anchors have been designed. The left one is the basic design which will be used as default. The middle one is used for tube layout shape adjustment when an anchor installation position error is greater than 2 cm but smaller than 7 cm. The tube layout shape is an important issue to control the friction, which will be discussed in details in Sec. \ref{section:sec-friction}.

\begin{figure}[htp]
    \centering
    \includegraphics[width=0.3\textwidth]{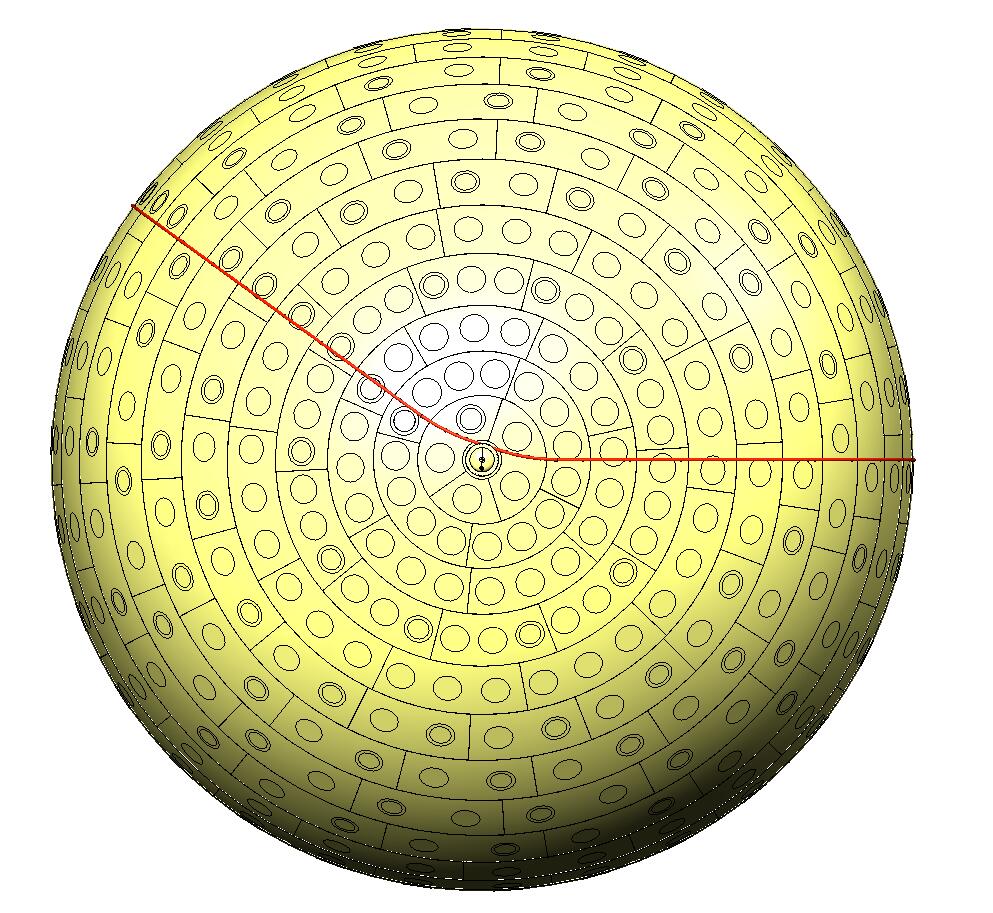}
	\caption{Snapshot of guide tube deployment, from bottom view. The yellow ball is the acrylic sphere with the LS inside. The circles on CD are the installation nodes for the supporting bars. The red line going between the nodes shows the route of GT.  (not to scale)}
	\label{fig:TubeRoute}
\end{figure}

The tube, attached on CD outer surface with the support of anchors, is deployed as shown in Fig.\ref{fig:TubeRoute}. Instead of a simple ring, the angle between two planes where the both semi-circles of the tube lie is 144\degree (at $\theta$ = 123.4\degree and 267.4\degree). This depends on the deployment of connection bars between the acrylic sphere and the stainless steel frame. 
%The both halves of the tube will be installed along the longitudinal lines at $\theta$ = 123.4\degree and 267.4\degree, respectively. 

The tube provides a running track for the radioactive source. During the experiment, 6 different sources will be used in GTCS as listed in Table \ref{tab:source}. Calibration with an Am-C source will be performed monthly, while the others will be used as the special calibration for the purpose of understanding the detector performance and simulation configuration. 
The source is enclosed in the source carrier, which will be inserted and guided by the tube and cable. To avoid losing source inside the tube, the structure of the source carrier is specially designed, and other safety measures are adopted. They will be introduced in details in section \ref{sec-security}.

\begin{table}[h]
\centering
\caption{\label{tab:source} Radiation Sources in GTCS.} 
\smallskip
\begin{tabular}{l|l|l|l}
\Xhline{1.2pt}
Source & Type & Energy (MeV)          & Use Mode  \\ \hline
$Am-C$    & $n,\gamma$  & 2.2 MeV   & Monthly       \\\hline
$^{40}K$  & $\gamma$    & 1.461            & Special Calibration \\\hline
$^{54}Mn$ & $\gamma$    & 0.835            & Special Calibration \\\hline
$^{60}Co$ & $\gamma$    & 1.17+1.33        & Special Calibration \\\hline
$^{68}Ge$ & $e^+$           & 0.511+0.511 & Special Calibration \\\hline
$^{137}Cs$& $\gamma$    & 0.662          & Special Calibration\\
\Xhline{1.2pt}
\end{tabular}
\end{table}

\subsection{The control system}
A control system guides the source to the designated positions.
To achieve this, two cables are tied on the both sides of the source carrier and are used to control the source position by adjusting their length simultaneously with the winding machines. Each of the cables is 150 m long to make sure the source position can cover the whole route. 

\begin{figure}[h]
    \centering
    \subfigure[]{    
		\label{fig_WMphoto}     
	    \includegraphics[height=0.25\textwidth]{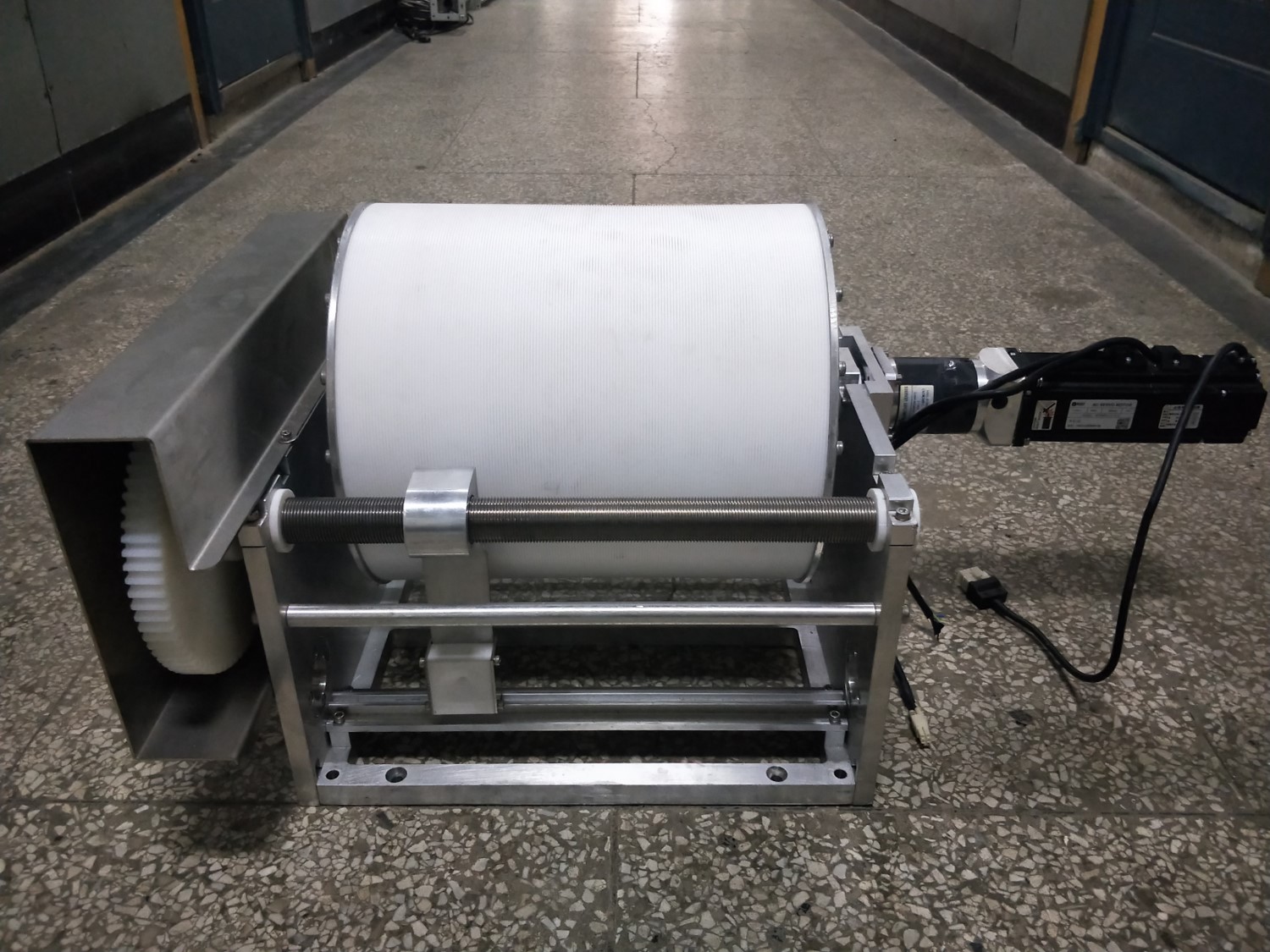}}
    \subfigure[]{
		\label{fig_WMDrawing}     
	    \includegraphics[height=0.25\textwidth]{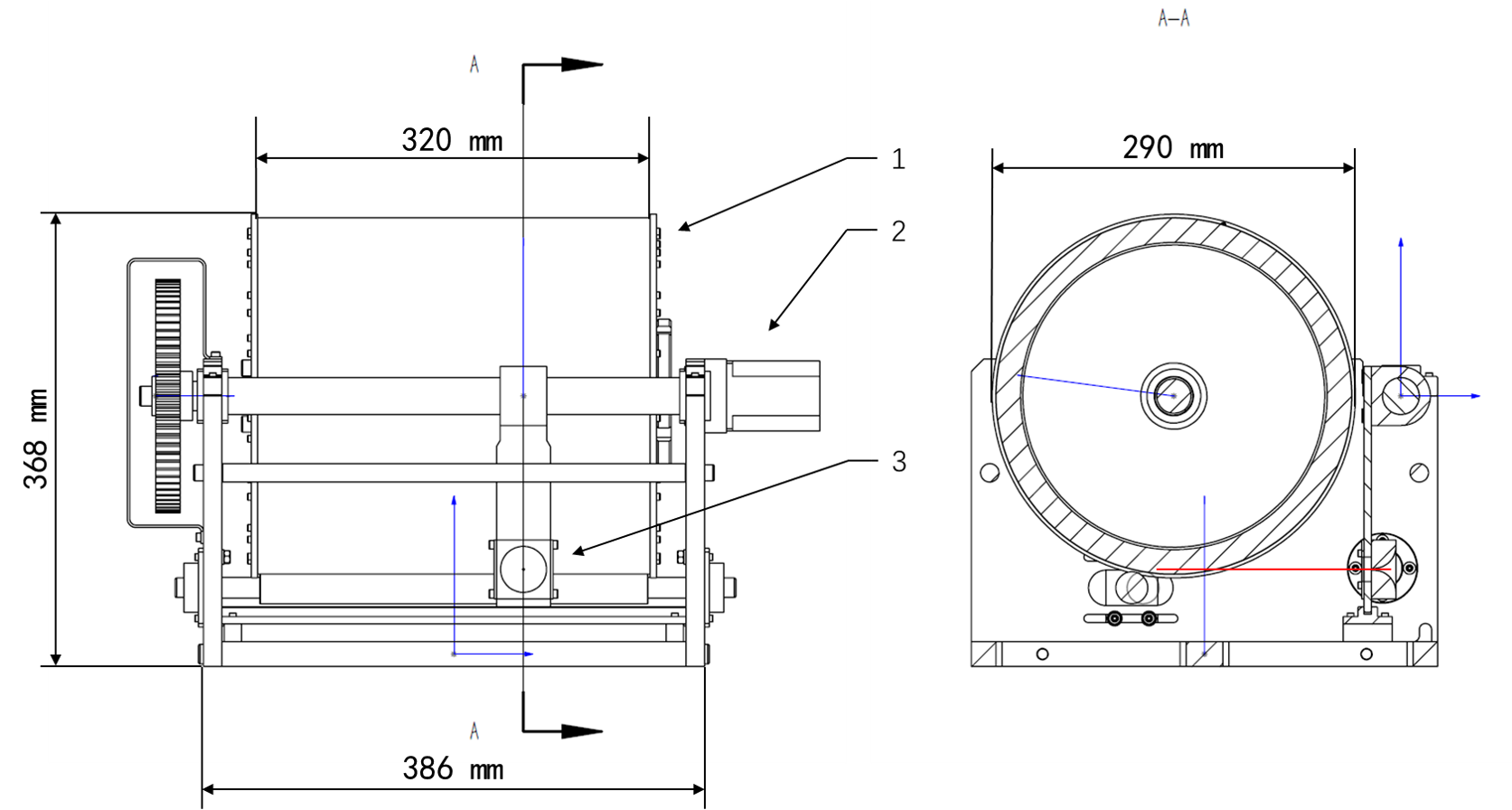}}
%    \subfigure[]{    
%		\label{fig_WMfront}     
%		\includegraphics[height=0.3\textwidth]{figure/WindingMachine3D.png}}
%    \subfigure[]{    
%		\label{fig_WMleft}     
%		\includegraphics[height=0.3\textwidth]{figure/WindingMachine3D-LeftCut.png}}
	\caption{Photo (a) and Scheme (b) of the winding machine. 1. spool, 2. motor, 3. cable position limit. The red line is the cable to be wound. With the help of the cable position limit, the cable can be wound on the correct position of the spool. One turn of the winding spool will release or retrieve \textasciitilde 0.91 m cable.}
	\label{fig:windingmachine}
\end{figure}

The winding machines and their auxiliary facilities are designed in order to manage such a long cable safely and effectively. The appearance of winding machine can be found in Fig.\ref{fig:windingmachine}. The cable will be wound on the PTFE spool. There is a 150 m long helical groove on the spool, with sectional depth and width of 1.5 mm. This groove holds the cable. With the help of the gears and the cable position limit, the cable can be wound on the spool circle by circle without error.

\begin{figure}[]
    \centering
    \subfigure[]{    
		\label{fig_TS}     
	    \includegraphics[height=0.3\textwidth]{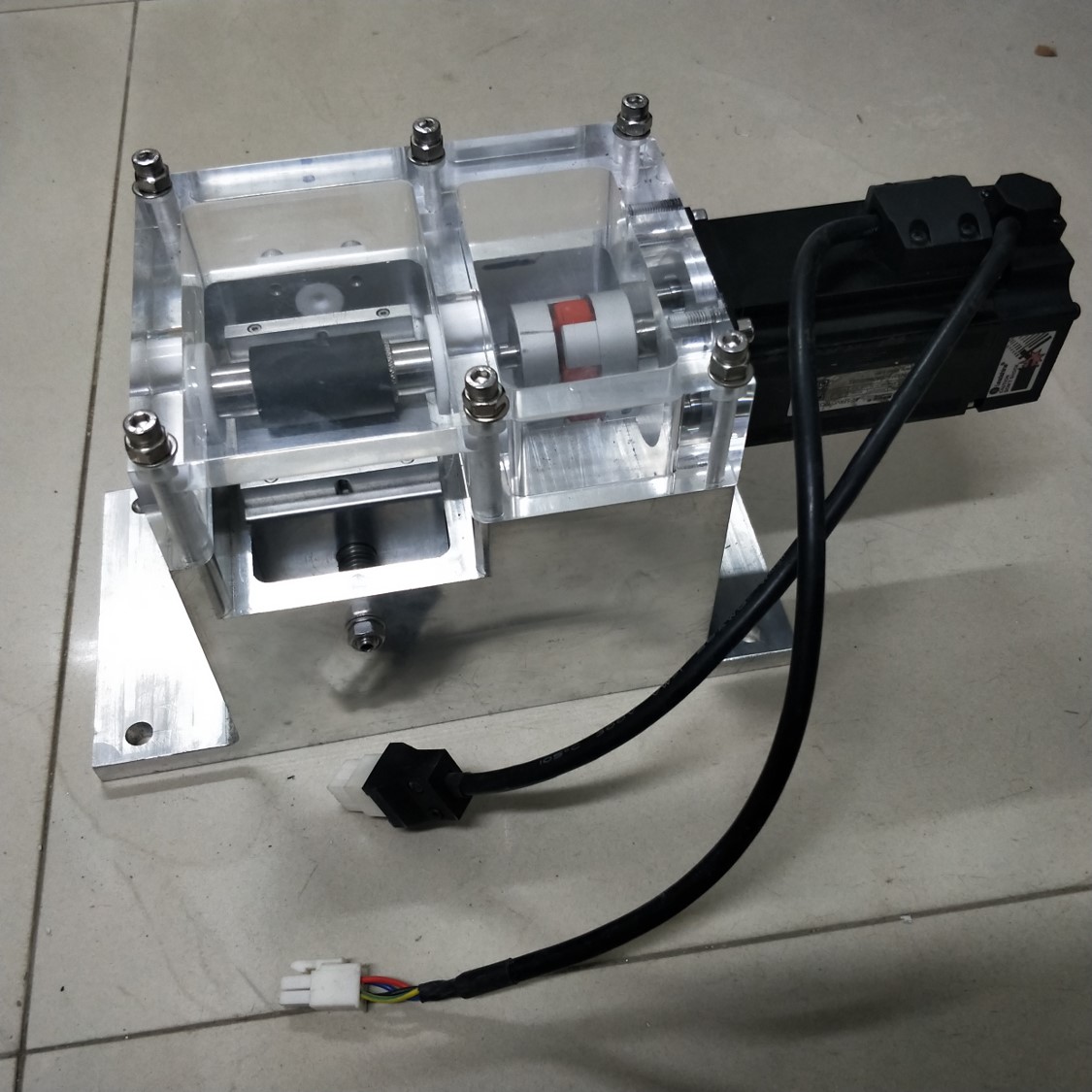}}
    \subfigure[]{
		\label{fig_TM}     
	    \includegraphics[height=0.3\textwidth]{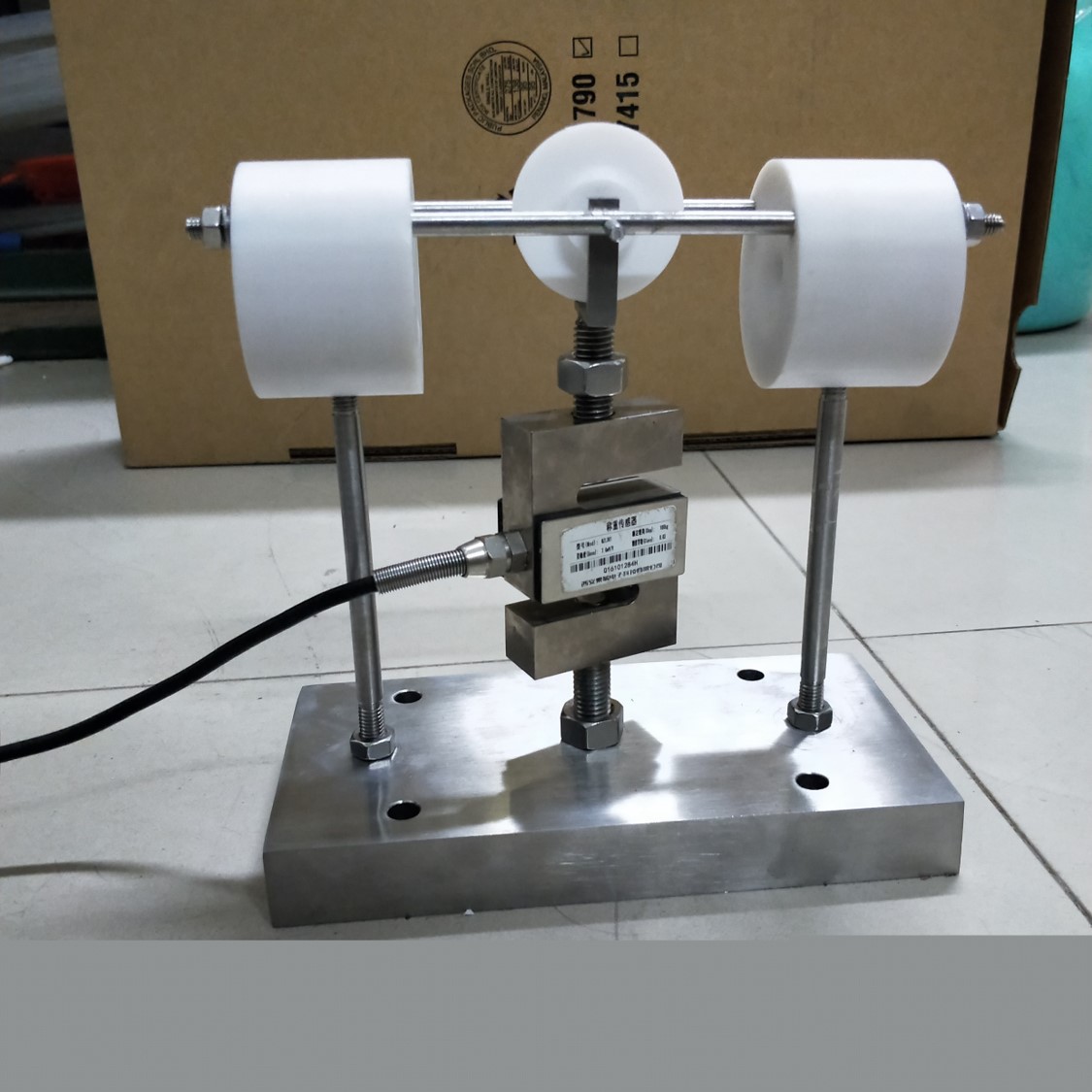}}
    \par
    \subfigure[]{
		\label{fig_PS}     
	    \includegraphics[height=0.26\textwidth]{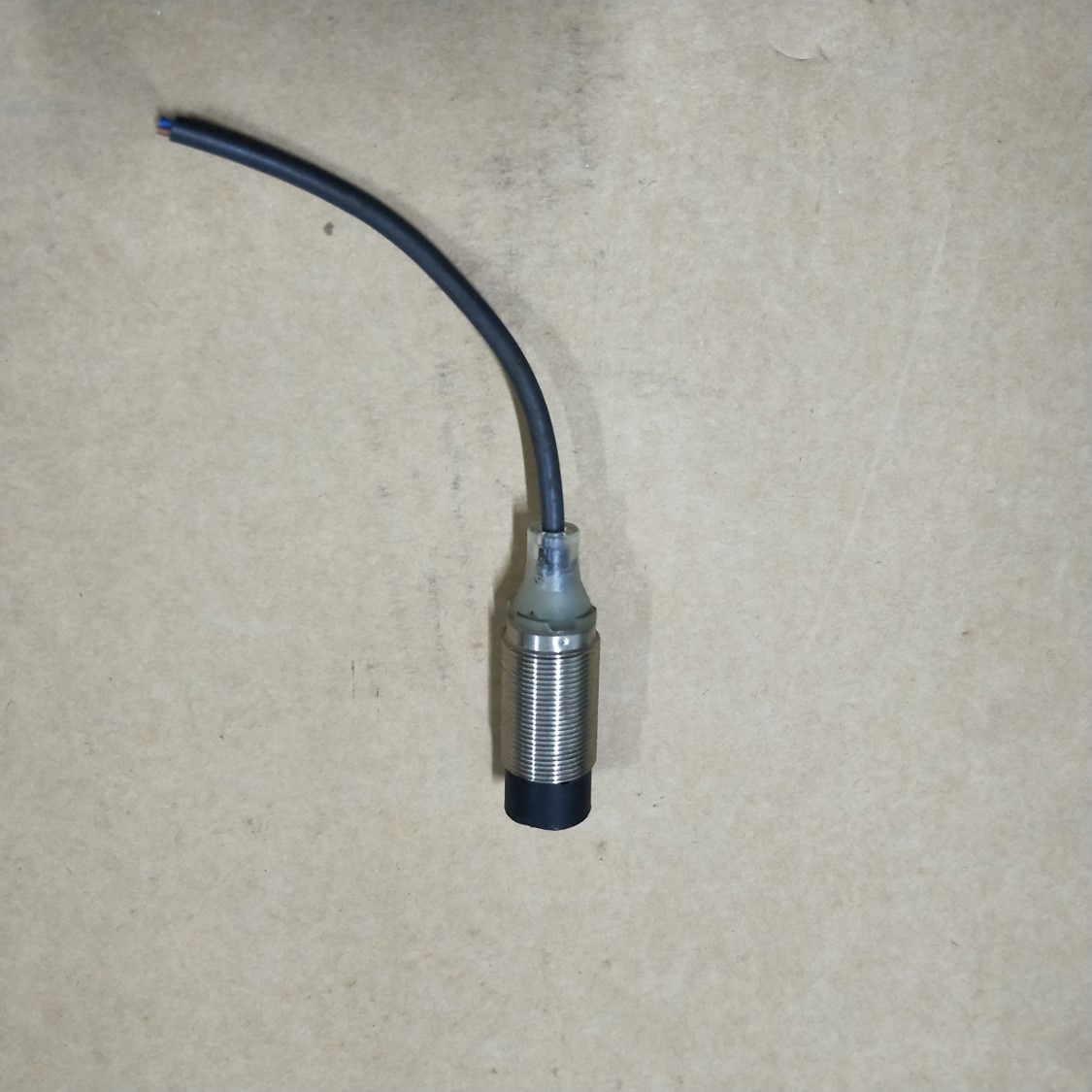}}
    \subfigure[]{
		\label{fig_SB}     
	    \includegraphics[height=0.26\textwidth]{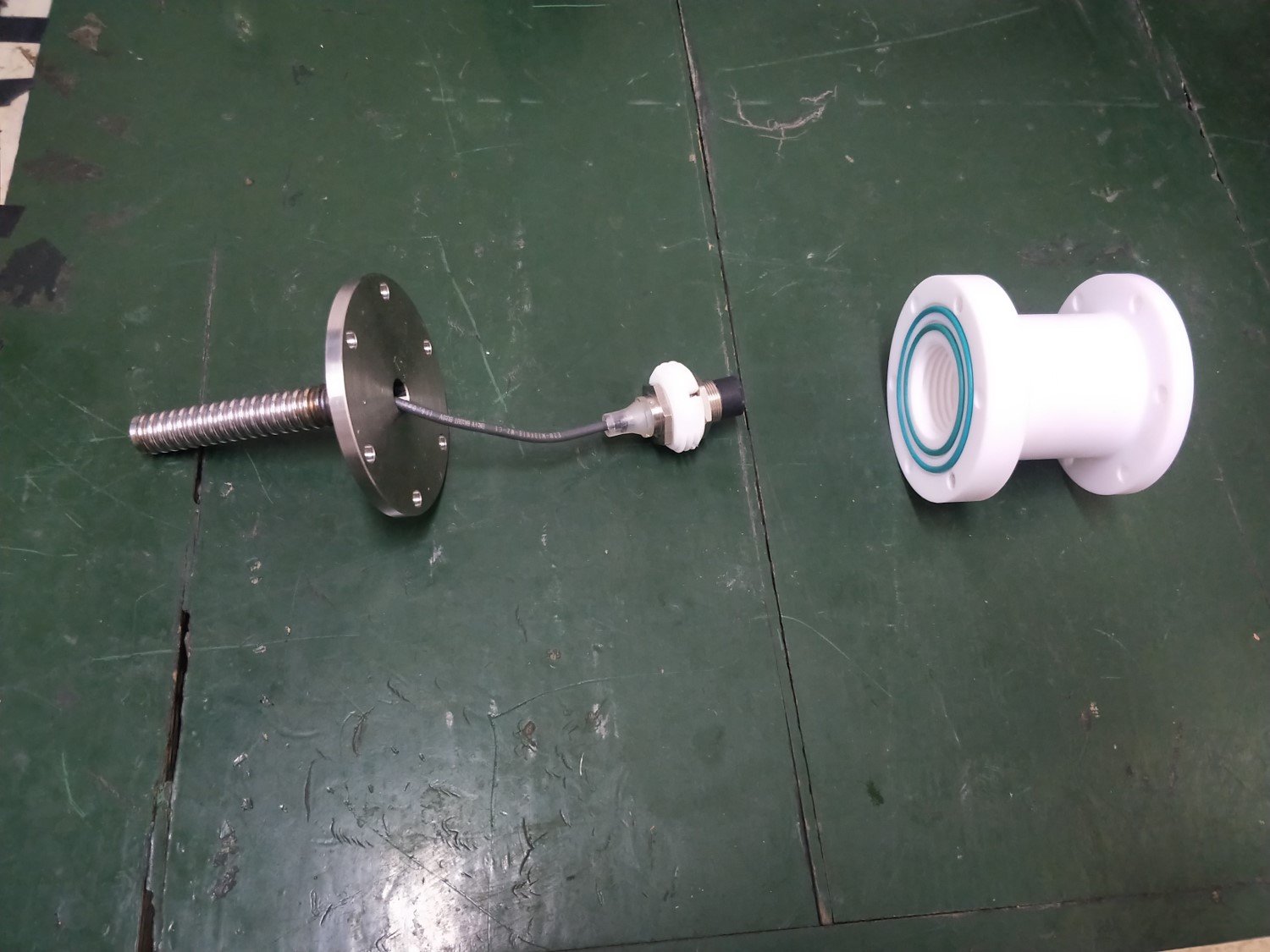}}
	\caption{Demonstration of the auxiliary facilities. (a) Tension Maintainer. (b) Tension Sensor. (c) Source Position Sensor. (d) Sealing Box of the Source Position Sensor}
	\label{fig:sensors}
\end{figure}

Each winding machine is driven by a ACM6004L2G servo motor and a L5-400motor driver,, produced by Leadshine Technology Co.,Ltd. The speed of motors can be controlled by computer via a PLC-based control board produced by Yokogawa Electric Corporation. \cite{L5Z} 

For the purpose of navigating the system safely and accurately, some auxiliary facilities are also designed in the control system including the tension maintainers, the tension sensors and the source position sensors. They are shown in Fig.\ref{fig:sensors} and will be introduced in section \ref{section:verify}. 

\iffalse
\begin{figure}[htp]
	\centering
    \subfigure[]{    
          \label{fig:motor}     
          \includegraphics[height=0.25\textwidth]{figure/motor.jpg}}
    \subfigure[]{    
          \label{fig:driver}     
          \includegraphics[height=0.25\textwidth]{figure/Driver.jpg}} 
    \subfigure[]{    
          \label{fig:PLC}     
          \includegraphics[height=0.25\textwidth]{figure/PLCboard.jpg}}
	\caption{Photo of (a) the motor on winding machine, (b) the driver of motor and (c) the PLC control board.}
	\label{fig:motioncontrol}
\end{figure}
\fi

\begin{figure}[htp]
	\centering
    \subfigure[]{    
          \label{fig_Box}     
          \includegraphics[height=0.6\textwidth]{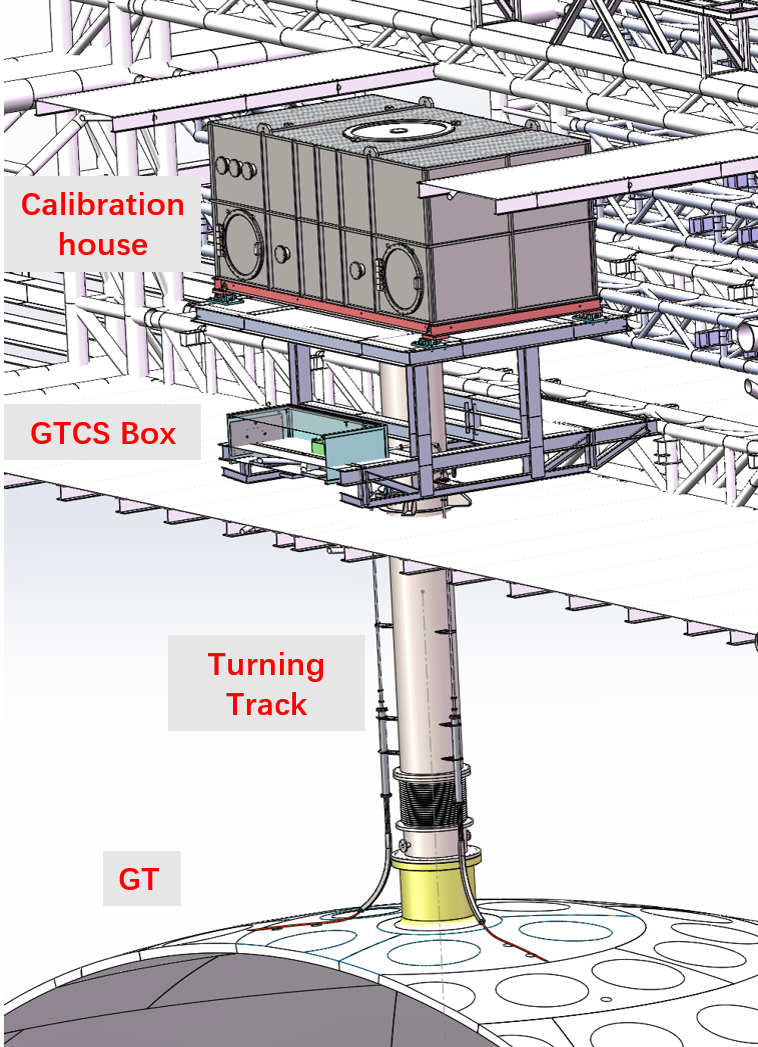}}
    \subfigure[]{    
          \label{fig_ConnectTube}     
          \includegraphics[height=0.6\textwidth]{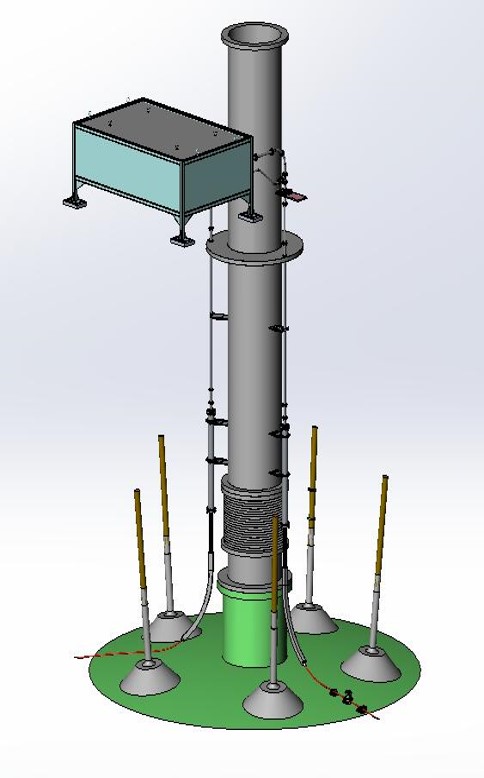}} 
	\caption{(a) General view of GTCS box and turning tracks. (b) The turning track is fixed on the chimney, guiding the PTFE tube to enter the GTCS box.}
	\label{fig:ConnectionTube}
\end{figure}

All these control parts are integrated in the GTCS box, which is made of stainless steel. The box will be installed on the lower floor of TT bridge. The GTCS box and the guide tube are connected by turning tracks. As shown in Fig.\ref{fig:ConnectionTube}, the turning tracks are a set of straight and curved stainless steel tubes near the Chimney and the GTCS box. They are assembled on the stainless steel part of Chimney with customized holders. The tube is well enclosed by the turning track and the GTCS box to avoid the light interfering into the CD.

\subsection{Remote navigation software}

A Programmable Logic Controller (PLC) is used to control the whole system. It will monitor the status of the system with the help of sensors and send commands to change the status manually or automatically. 
In order to monitor and navigate the system with the PLC remotely, a control software has been programmed based on the ladder logic language.

\begin{figure*}[h]
	\centering
    \subfigure[]{    
          \label{fig_soft}     
          \includegraphics[width=0.9\textwidth]{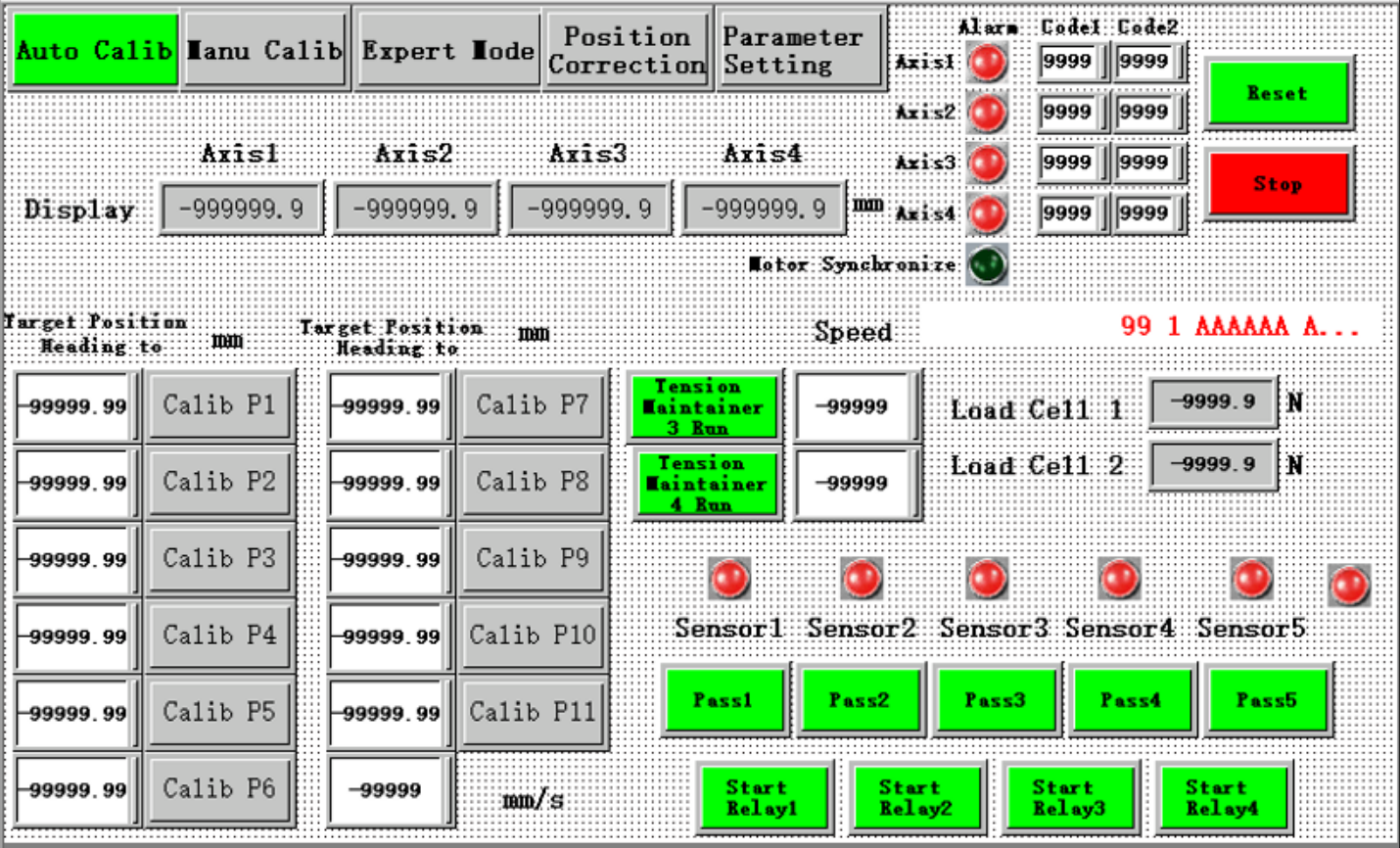}}
    \subfigure[]{    
          \label{fig_logical}     
          \includegraphics[width=0.9\textwidth]{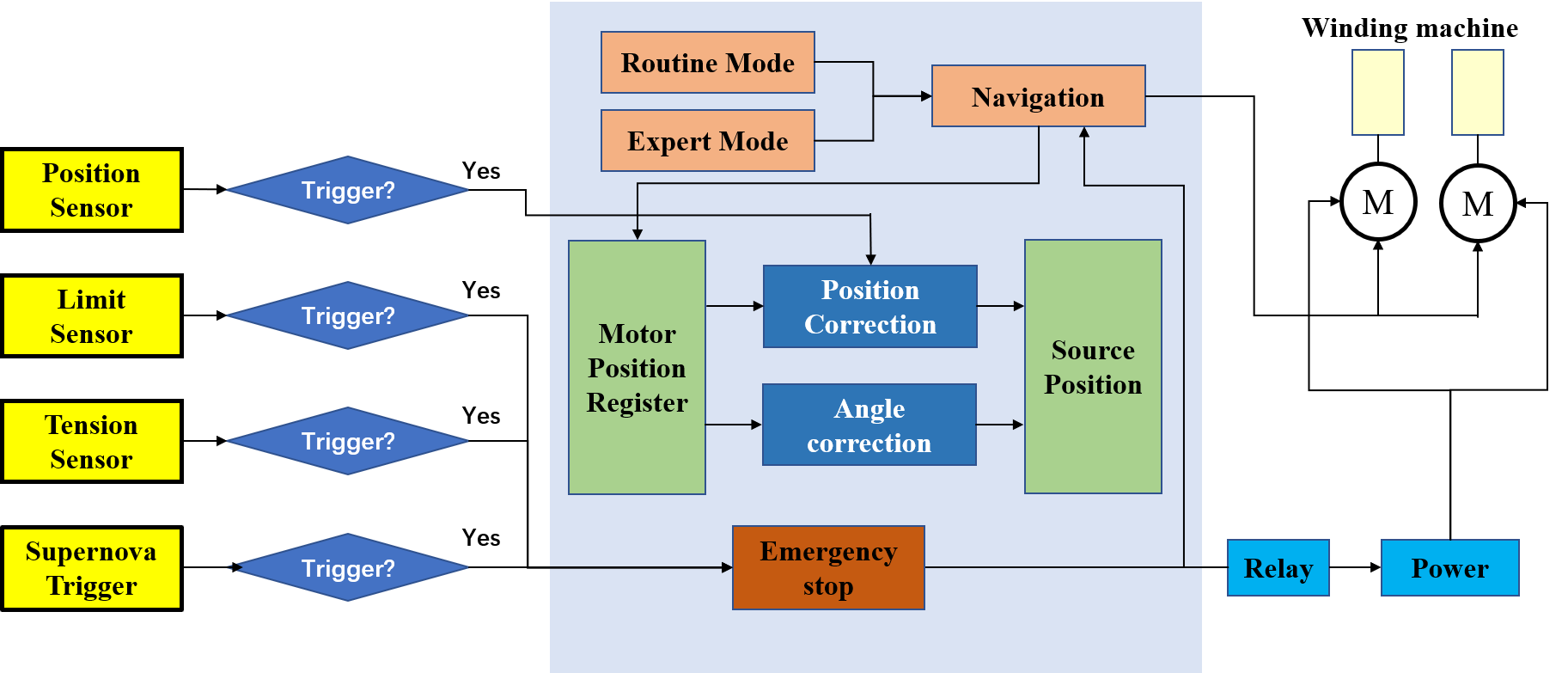}} 
	\caption{(a) GUI of the remote navigation software. (b) Flow chart of GT control software.}
	\label{fig:software}
\end{figure*}

As shown in Fig.\ref{fig:software}, there are several functions embedded in advance. Operator can navigate the system in routine mode, under which the distance and speed are pre-defined. The software will automatically correct the source position according to the signal from position sensors, as well as the angle bias caused by the motion of cable position limit. The software sets an emergency stop when any of the limit sensors and tension sensors is triggered. Meanwhile, it also provides the possibility of emergency stop when any supernova trigger signal is received. The software could also work in expert mode, under which the motors can run individually, the speed can be set freely, and the position sensors can be disabled.  
All the functions are embedded in the memory of PLC control board instead of the computer, so the system can be stopped in emergency without delay. The data would be uploaded to the computer for monitoring and storing synchronously. 

\subsection{Radioactive background estimation}

In order to reach the physical goal as soon as possible, the radioactivity background of JUNO is required less than 10 Hz within fiducial volume when the energy deposition ($E_{dep}$) is more than 0.7 MeV. \cite{a,JieZhao}
We aim to limit the background contribution of GTCS under 100 mHz. In this section, we are about to make a background rate estimate based on the simulation and radioactive measurement of the relevant components. 

The potential background of GTCS is mainly from the radioactivity of the components attached on the CD, including the tube, the cable and the positioning sensors. 
Their radioactivities have been measured with the HPGe detector in Jinping laboratory \cite{CJPL}.
The geometrical acceptance factors are applied to describe the fiducial volume selection and the energy deposition selection efficiency, which is estimated via simulation in SNIPER. \cite{SNIPER}.
The geometrical acceptance factors are the ratio of the remained event counts after selection by 1) $E_{dep}$ > 0.7 MeV and 2) radius of reconstructed $E_{dep}$ center < 17.2 m. The result can be found in Table \ref{tab:BKG}.

\begin{table}[h]
\centering
\caption{\label{tab:BKG} Background level estimation of GTCS. The radioactivities and the total background levels are provided with the upper 90 \% CL limit on the estimated rate. The errors of the geometrical acceptance factors are the statistic errors. The background has been scaled to the total amount to be used. The masses of tube and sensors to be used on CD surface are 20 kg and 341 g, and the length of cable on CD surface is 111 m.}
\smallskip
\begin{tabular}{c|c|ccc}
\Xhline{1.2pt}
         & Isotope & Tube      & Cable & Sensor \\ \Xhline{1.2pt}
\multirow{4}{*}{\tabincell{c}{Radioactivity \\(mBq)}} 
         &$ ^{40}K    $     & <2452      & <184  & <5563   \\
         &$ ^{60}Co  $      & <81.0        & <147  & <22.38   \\
         &$ ^{232}Th $     & <0.25        & <31.7   & <3936   \\
         &$ ^{238}U  $     & <0.40       & <255  & <4010   \\
\hline
\multirow{4}{*}{\tabincell{c}{Geometrical \\Acceptance \\Factors (\%)}}
     & $^{40}K$     & $0.104\pm0.007$  & $0.099\pm0.007$ & $0.072\pm0.003$  \\
     & $^{60}Co$   & $1.235\pm0.025$  & $1.177\pm0.024$ & $1.072\pm0.011$  \\
     & $^{232}Th$   & $0.145\pm0.009$  & $0.127\pm0.008$ & $0.111\pm0.004$  \\
     & $^{238}U$   & $0.051\pm0.005$  & $0.054\pm0.005$ & $0.044\pm0.002$  \\ \hline
%\multicolumn{2}{c|}{\multirow{2}{*}{Total (mHz)}} & $3.56\pm0.17$ & $4.24\pm0.06$ & $72.64\pm0.60$ \\ \cline{3-5}
\multicolumn{2}{c|}{\multirow{2}{*}{Background (mHz)}} & <3.80 & <4.57 & <76.31 \\ \cline{3-5}
\multicolumn{2}{c|}{}                            & \multicolumn{3}{c}{<84.68} \\ \cline{3-5} 
\Xhline{1.2pt}
\end{tabular}
\end{table}

Then the estimated background from each structure is calculated according Eq. \eqref{eq:bkg}:

\begin{equation}
\label{eq:bkg}
B= \sum_i A_i \epsilon_i k_i
\end{equation}

where $\epsilon_i$ is the geometrical acceptance factors of the $i^{th}$ isotopes, which is chosen from the isotopes listed. $A_i$ and $k_i$ are the radioactivity and the number of gammas associated with each mother nuclide decay chain, respectively. $k$ of $^{60}Co$, $^{40}K$, $^{238}U$ and $^{232}Th$ are 1, 1, 14 and 10 respectively. The total background level is smaller than 84.68 mHz (90$\%$ C.L.), which meets our goal.

\section{Verification test}\label{section:verify}

Realistic tests were performed with a full size prototype to verify the designed functions and will be introduced in this section. The following issues will be addressed: 1) the measurement and control of the friction between the cable and tube. 2) the estimation of the source position uncertainty. 3) some enhancement measurements to make the system more reliable.

\subsection{Full size prototype}

\begin{figure*}[htp]
	\centering
    \subfigure[]{    
          \label{fig_general}     
          \includegraphics[width=0.4\textwidth]{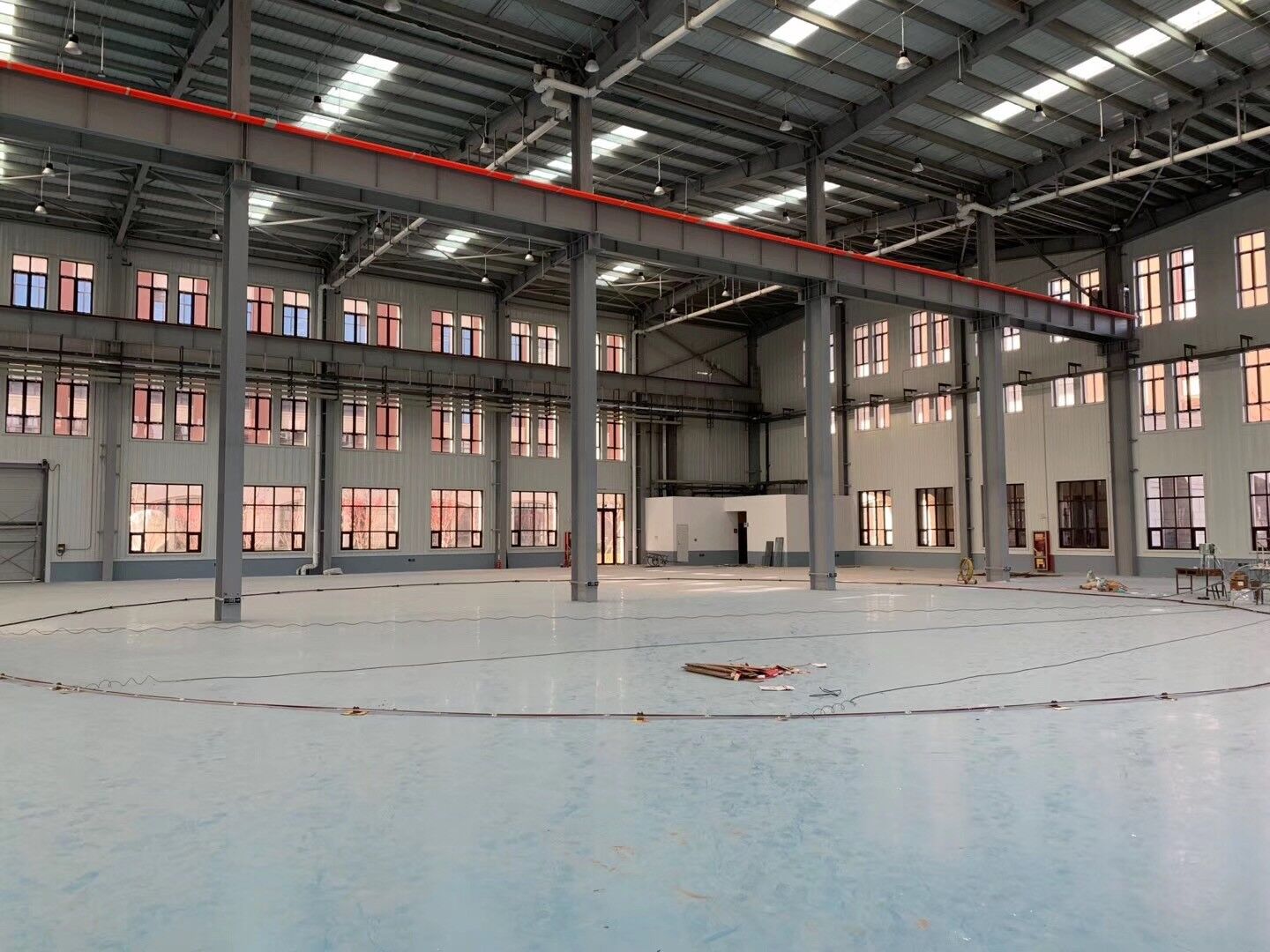}}
    \subfigure[]{    
          \label{fig_wind}     
          \includegraphics[width=0.4\textwidth]{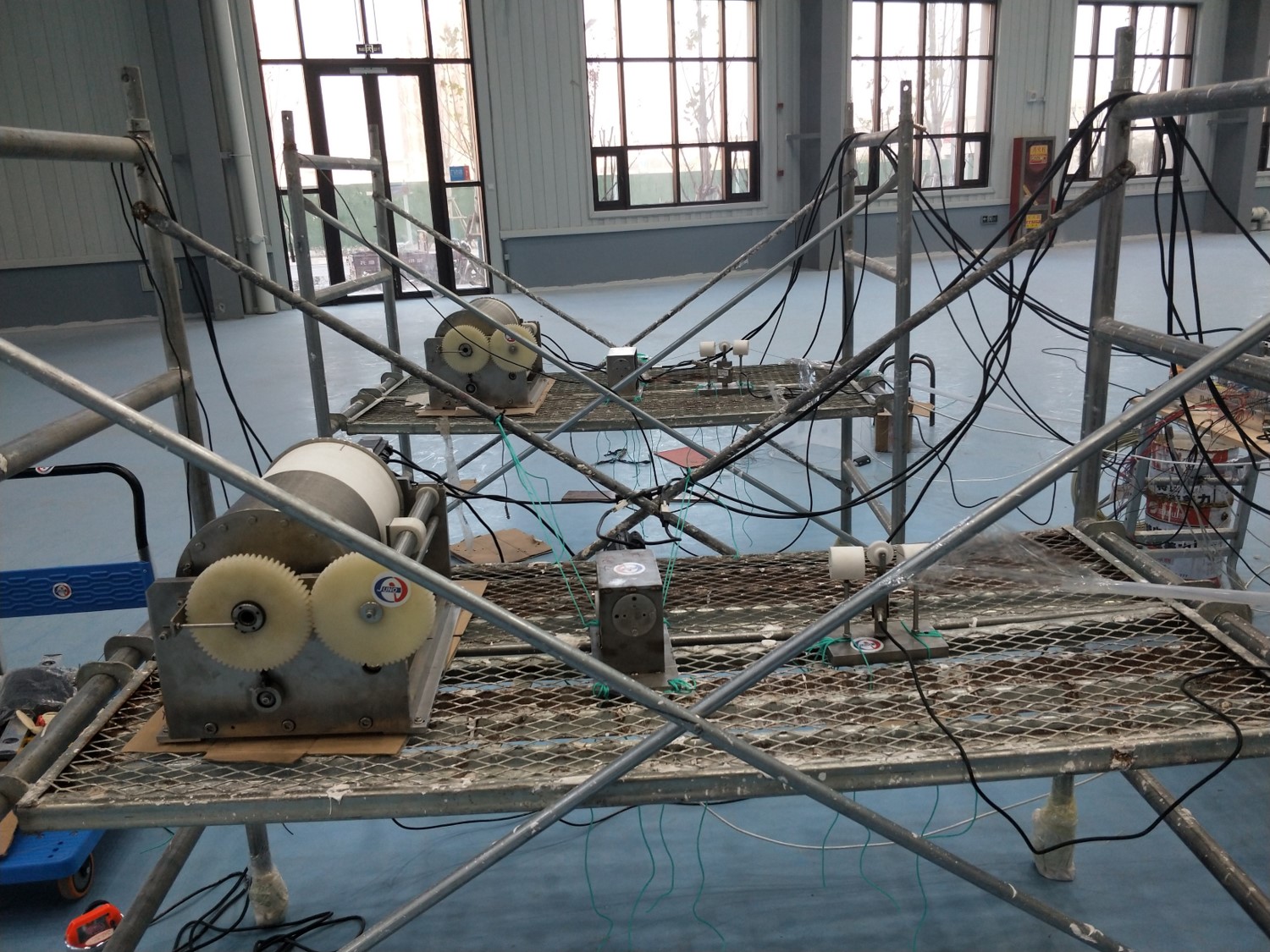}} 
	\caption{Layout of the verification experiment. (a) General layout. A horizontal track and tube with radius of 35.4m. (b) The layout of winding machine, tension maintainer and tension sensor.}
	\label{fig:Prototype}
\end{figure*}

The verification test is based on the full-size prototype of GTCS. 
A circular track with a diameter of 35.4 m was constructed horizontally on the ground. 
A tube was fixed on the track with cable ties and led to the winding machines. 
The winding machine, the tension maintainer, the tension sensors and the both ends of the tube were fixed on a 1 meter high table, so the tube can be filled with water as in the final experiment. 
The cable was threaded in the tube and then wound by the winding machines after going through the tension sensor and a tension maintainer. 
The system was controlled remotely on the computer with the navigate software. 
A layout of the prototype can be found in Fig. \ref{fig:Prototype}. 

The full size prototype was designed to test the synchronous run of GTCS in a similar environment to the final experiment. One of the differences was that the track was horizontal instead of vertical. The cable and the source carrier were not heavy enough to influence the result.  Another difference was the shape of the tube. 
Comparing to the final experiment, the 144\degree \ turning of tube at the bottom of CD was not mimicked. However, lifting the tube from the ground to the table introduced an extra 180\degree \ turning angle to the tube. So the turning angle was about 36\degree \  larger than that in final experiment. Therefore, as introduced in \ref{section:sec-friction}, the friction between the tube and cable should be larger than that in final experiment. Since we want the friction to be small, this difference won't risk the reliability of the full size prototype test results. 

Many basic functions of the system have been tested to make sure they can run synchronously. 
Essential parameters were also measured with the prototype. 
The test result are listed in Table \ref{tab:Verification}.

\begin{table}[hbp]
\centering
\caption{\label{tab:Verification} Data sheets of Verification Experiment. }
\smallskip
\begin{tabular}{m{7cm}|m{7cm}}
\Xhline{1.2pt}
Test Item                                                              &Result\\
\hline
Performance of the Winding Machine                  &  The cable can be released and retrieved well with more than 50 cycles.\\
\hline
Performance of the Tension Maintainer               &  Tension on winding spools were maintained over 50 full positioning cycles. \\
\hline
Performance of the Tension Sensor                   &The position sensors could be used to reproducibly stop the cable movement within the desired position tolerance. \\
\hline
Performance of the Position Sensor              &The sensors can help to correct the sensor position. And the response of the sensors were immediate. \\
\hline
Friction between tube and cable                       &The friction is < 20N when the density of anchors is about 1 anchor / 2 m.\\
\hline
Force on anchor from tube                               &<10N, acceptable\\
\hline
Performance of cable threading                            &Successful Operation at 130 m long tube\\
\hline
Max friction between tube and threading fiber & 150N, acceptable\\
\hline
Coordinated run of the system                       & The integrated system worked well in the whole test. All the designed functions of the system were tested and achieved. Accidents like the cable breakage or source loss were avoided.\\
\Xhline{1.2pt}
\end{tabular}
\end{table}

\subsection{The friction investigation}\label{section:sec-friction}

One of the important issues in GTCS is the friction between the tube and cable. If the friction is too large, the reliability of the whole system will become worse. In the ideal case, if the tube is a perfect circle, the tension on each side can be described by the Euler-Eytelwein formula as shown in Eq \eqref{eq:tension} \cite{EularEquation}.

\begin{equation}
\label{eq:tension}
T_{1}=T_{0}e^{\mu \theta}
\end{equation}

Where $T_{1}$ is the tension on the retrieving side where the cable is getting out of the tube and retrieved by the winding machine. $T_{0}$ is the tension on the releasing side, which is opposite to the retrieving side. The retrieving and releasing sides are defined according to the source moving direction. $\mu$ is the friction coefficient between the tube and cable, and $\theta$ is the turning angle. Then, the tension $f$ would be

\begin{equation}
\label{eq:friction}
f=T_{1}-T_{0}=T_{0}(e^{\mu \theta}-1)
\end{equation}

Eq \eqref{eq:friction} has shown two ways to maintain a low friction between the cable and tube.
The first way is to make the tension on the cable in the releasing side as small as possible, because the friction and the tension on the cable in the retrieving side are proportional to it. Therefore, the tension on the cable on the releasing side has been required to be zero when the system runs.

Secondly, the exponential dependency on $\theta$ indicates that the turning angle of cable should be as small as possible. This will require the tube to be straight along its deployment direction. The shape of the tube depends on the installation error of the acrylic sphere and GT anchors. Although the anchors should have been installed along a longitudinal line, the actual anchor position biases are inevitable during installation, making the shape of tube unpredictable. 
In this test, we found that the friction between the cable and tube didn't change obviously if the anchor position bias is controlled within 2 cm. However, the maximum of acrylic sphere installation error is estimated to be larger than this value. So we optimized the design of anchors to fit this tolerance. As shown in Fig. \ref{fig:GTanchor}, the holder is designed with seven 2 cm holes to reduce the tube installation error caused by the error of GT anchor installation. If one of the anchors is biased from the benchmark line, the tube will go through the best fit hole. In this way, the bias can be smaller than 2 cm.

\subsection{Cable threading method}

\begin{figure*}[h]
    \centering
    \subfigure[]{    
          \label{fig_Threada}     
          \includegraphics[height=0.35\textwidth]{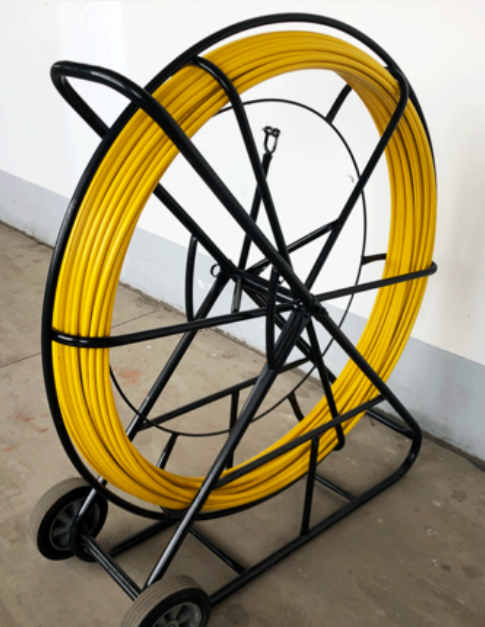}}
    \subfigure[]{    
          \label{fig_Threadb}     
          \includegraphics[height=0.35\textwidth]{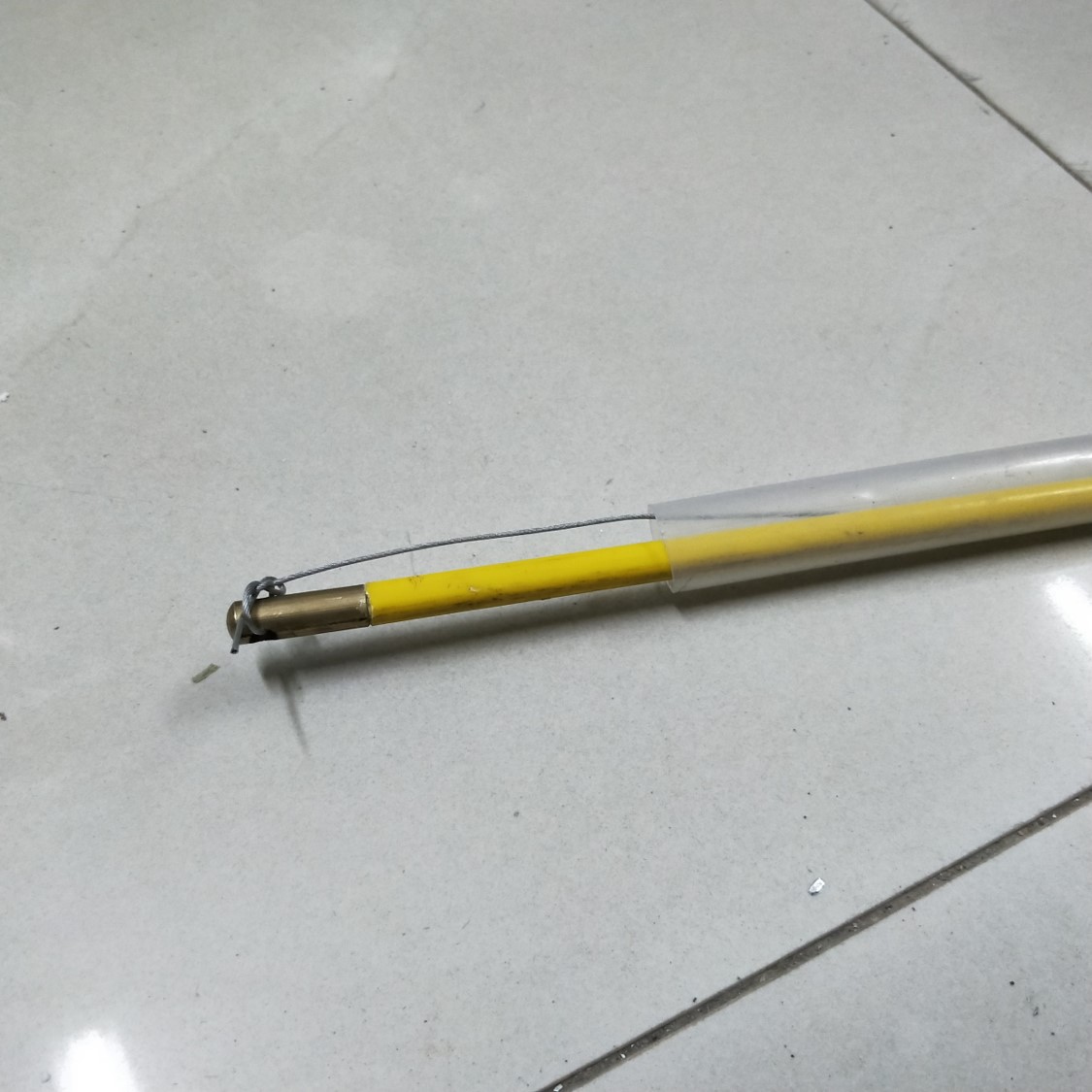}}
	\caption{Photos of the threading fiber. It is widely used in the industry.}
	\label{fig:ThreadingMachine}
\end{figure*}

Another issue related to the friction is how to thread the cable into a 120 m long curved tube, especially when the tube is long and thin and the cable is soft. 
The problem will become much more difficult when the threading is required to be processed when the GT have been installed on the CD, because the GTCS is designed to have the chance to be resumed even if the cable breaks after everything is installed. The common threading method for electricians, as shown in Fig.\ref{fig:ThreadingMachine}, has been applied and tested on the prototype. A 120 m long Epoxy Fiber Glass threading fiber is used to carry the cable through the tube. However, a direct attempt of cable threading failed because of the large friction in the final stage. In order to reduce the friction, a PTFE tape was coated on the threading fiber as a skin. This method will be the basic way to thread the cable because it can thread the whole route easily by a single person.

\subsection{Source positioning uncertainty}

The uncertainty of source positioning is quite important because the response along GT is position dependent. In reference \cite{GT_1st}, simulation works indicated a maximum GT response bias of 0.15\% corresponding to the positioning uncertainty of 10 cm. We hope to control it within 10 cm. Therefore, a verification to the source position uncertainty on the prototype is necessary. 

The uncertainty of source positioning usually comes from: 1) the deformation of the acrylic sphere. 2) the installation error of the GT anchor. 3) the deformation of the cable. The first two are unpredictable but will not change once the experiment installation is finished. Meanwhile, the third one depends on the tension on the cable. 
%So we can make an estimation on the prototype. Corresponding methods have been made and introduced as following.

We designed the position sensors to correct the source position uncertainty caused by the acrylic sphere deformation. They are metal sensors which can be triggered by the metal in the source carrier. 6 sensors are deployed along the calibration route with the GT anchors. The installed positions of these anchors will be surveyed. The feedback signal when the source is passing can be used to correct the position of source with linear interpolation. 

The installation error of GTCS anchors will also increase the positioning uncertainty. Considering the adjustable GT anchor, the maximum installation error of tube in horizontal direction is 2 cm. The positioning bias caused by the anchor installation error should be the difference of the tube length and the anchor distance in latitude direction. In the extreme conditions when two neighbor anchors are 2 cm biased reversely in horizontal direction, assuming the tube is straight, positioning bias along the tube would be 0.04 cm. 
Usually, we suppose only few anchors would be installed incorrectly, so this kind of errors should be very small and ignorable. 

\begin{figure*}[htp]
    \centering
    \includegraphics[width=0.6\textwidth]{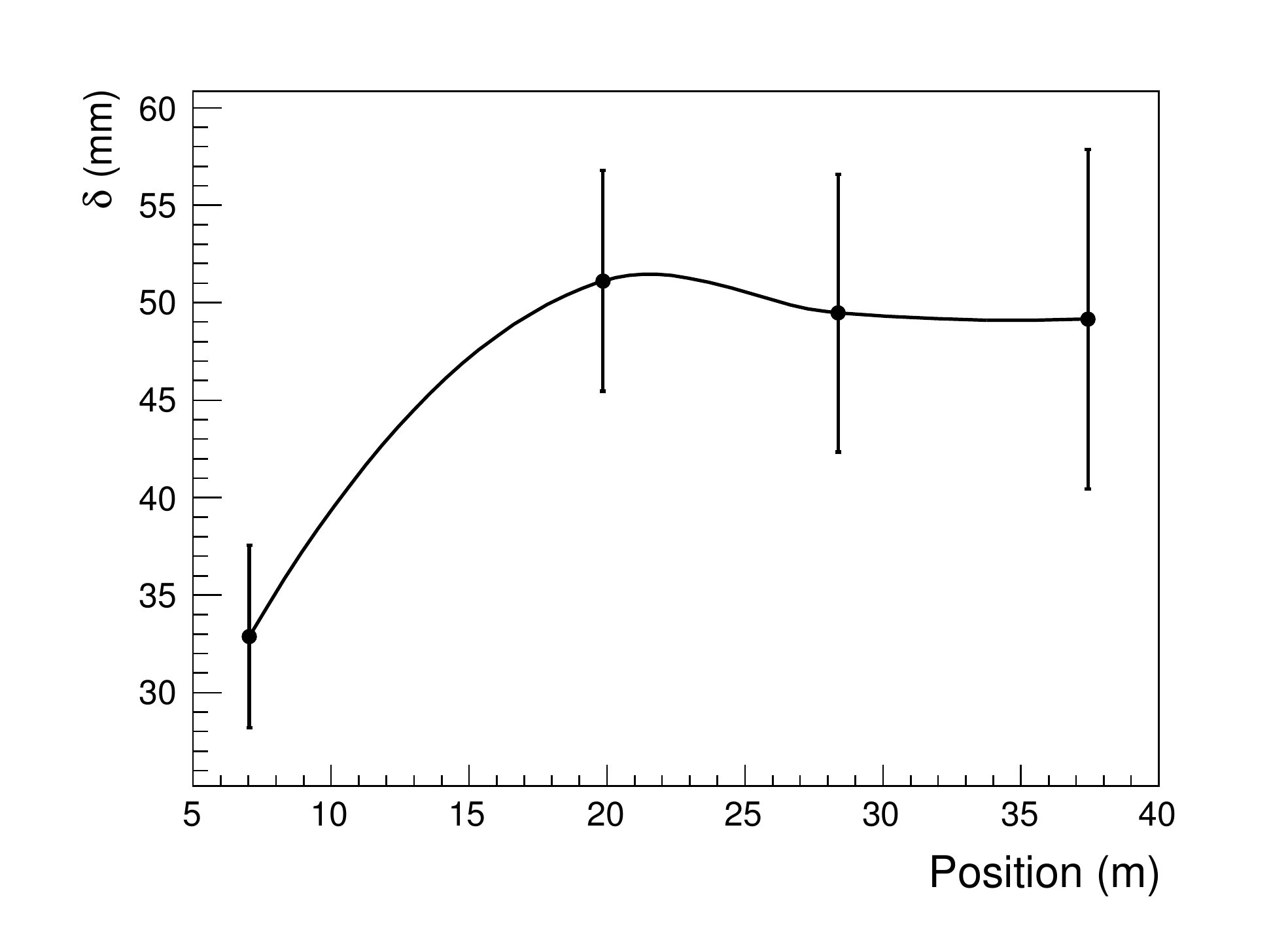}
	\caption{Cable deformation in length at the specified positions in the prototype test. The x-axis is the position of the measured points where the position sensors were installed.}
	\label{fig:positionerrorverify}
\end{figure*}

The cable deformation is another consideration. We are able to make an estimation on it with the help of position sensors. In the prototype experiment, we deployed 4 sensors along the tube and their positions are measured. Then, we navigated the source to approach the sensors. The maximum tension on the cable was controlled to be around 20 N. The motors stopped immediately when the position sensors were triggered. However, the source would continue to run shortly as the cable tension was released. Therefore, the delayed short running distance would approximately be the source position error caused by the cable deformation. We repeated this procedure 10 times and the results are shown in Fig. \ref{fig:positionerrorverify}. From this plot, the positioning uncertainty from cable deformation should be smaller than 6 cm. 

As a result, the positioning error has been verified to be smaller than 10 cm, which meets our requirement.

\subsection{Reliability considerations and risk control} \label{sec-security}

In the running strategy of GTCS, keeping the source under control is critical. In particular, it is mandatory that the source can be retrieved from the tube in any case. Two principles have to be taken into account concerning this. Firstly, the tension on the cable should be small. Secondly, the source carrier should be reliable. 
The cable has been tested to endure a maximum tension of 800 N, but the joint point of the cable and the source carrier can only hold a maximum tension around 150 N. So the maximum tension on cable should be controlled smaller than 50 N for safety considerations. 

\begin{figure*}[htp]
	\centering
    \subfigure[]{    
          \label{fig_TSb}     
          \includegraphics[height=0.35\textwidth]{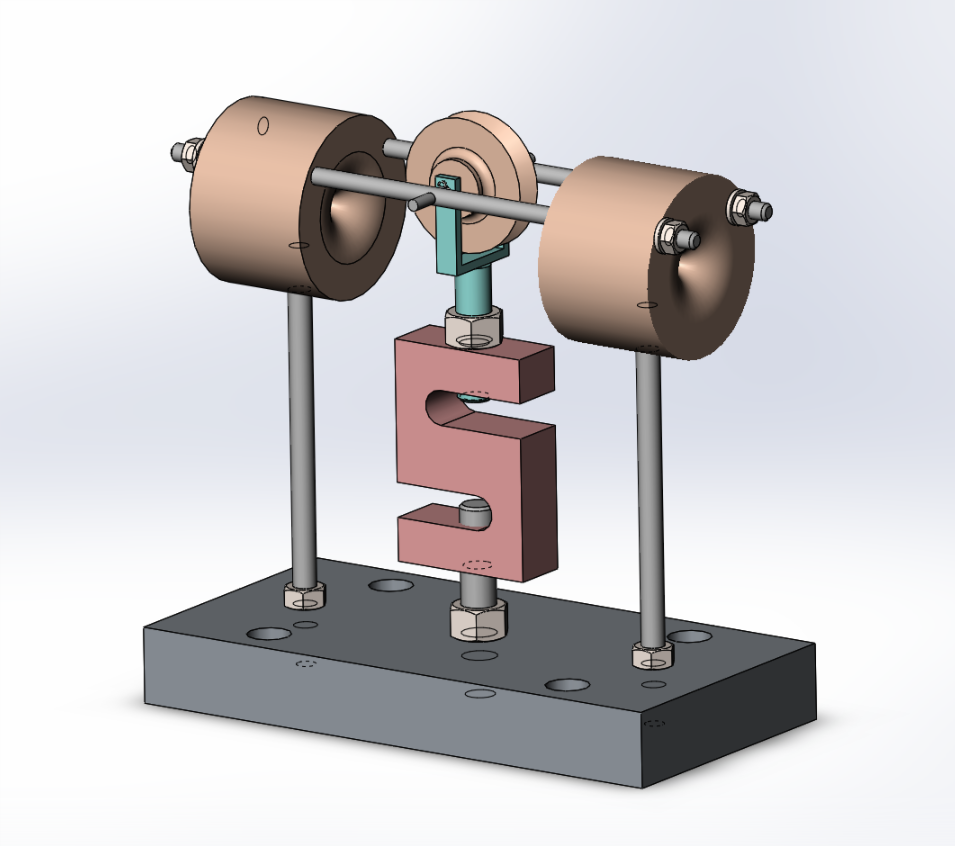}}
    \subfigure[]{    
          \label{fig_TSc}     
          \includegraphics[height=0.35\textwidth]{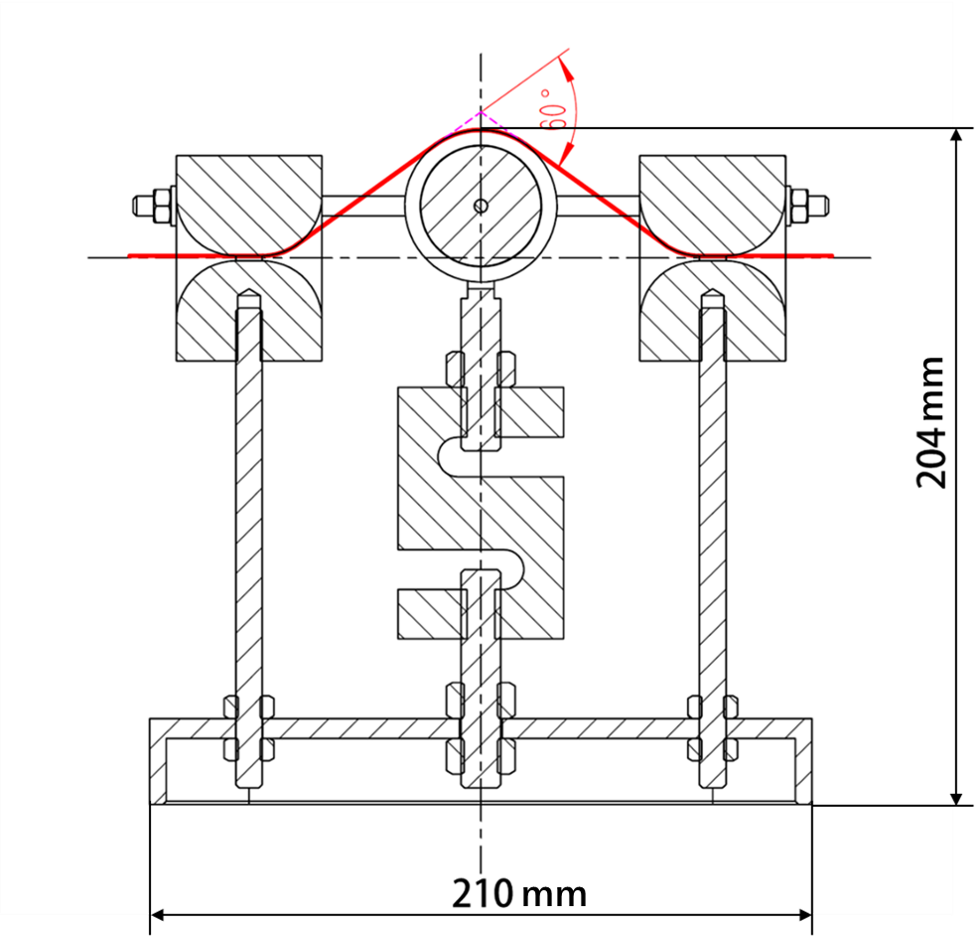}} 
	\caption{3D demonstration (a)  and profile drawing (b) of tension sensor. The angle between two cables are 120 \degree . }
	\label{fig:tensionsensor}
\end{figure*}

To keep the cable safe, its tension should be monitored. Two tension sensors have been deployed on both sides where the maximum tension could appears. As demonstrated in Fig. \ref{fig:tensionsensor}, the cable will go through a pulley and the holes of the limit blocks with an angle of $120^{o}$. Once the tension is found larger than 50 N, the winding progress will stop automatically for protection. 

\begin{figure*}[htp]
	\centering  
    \subfigure[]{    
          \label{fig_SCa}     
          \includegraphics[height=0.35\textwidth]{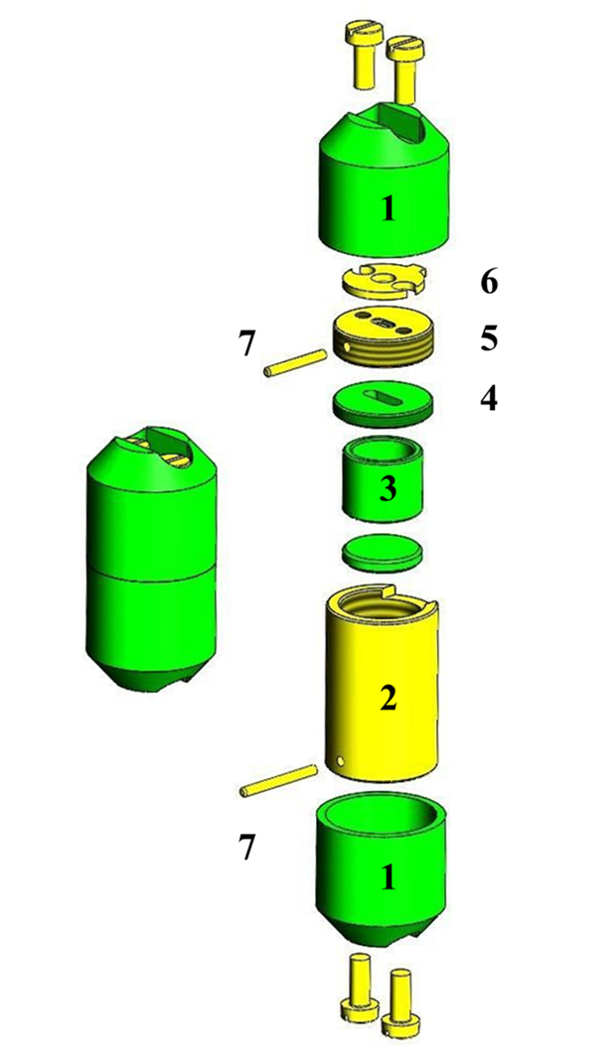}}
    \subfigure[]{    
          \label{fig_SCb}     
          \includegraphics[height=0.35\textwidth]{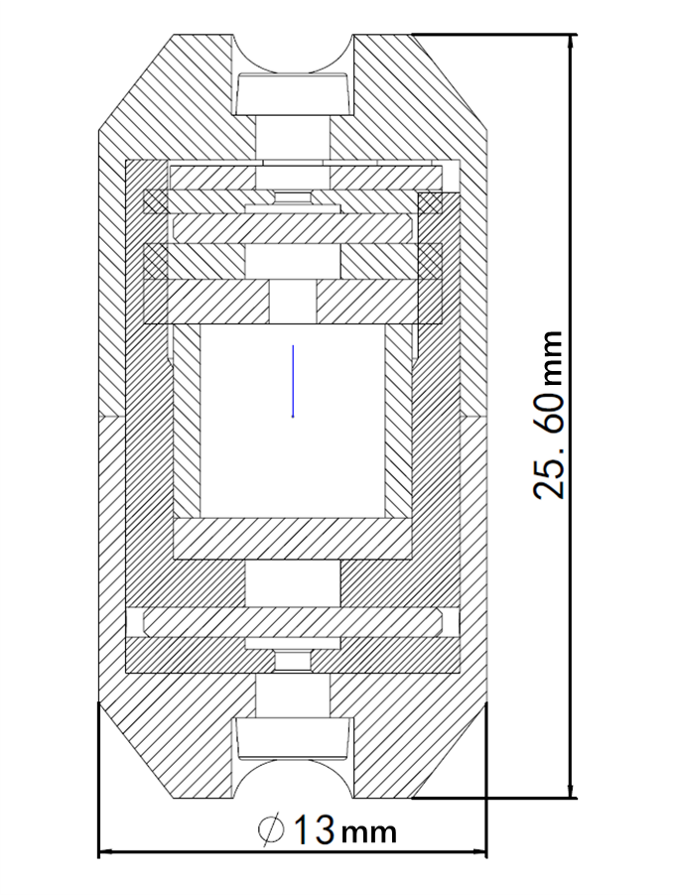}}
	\caption{Design of the source carrier. 1. outer shell. 2. inner shell. 3.  source holder. 4. PTFE plug. 5. stainless steel plug. 6. retaining key. 7. cable dowel, the cable is tied on it to drag the source carrier. PTFE and stainless steel components are in green and yellow. }
	\label{fig:source}
\end{figure*}

\begin{figure*}[h]
	\centering
%    \subfigure[]{    
%          \label{fig_TMa}     
%          \includegraphics[height=0.6\textwidth]{figure/TensionMaintainer2.jpg}}
    \subfigure[]{    
          \label{fig_TMb}     
          \includegraphics[height=0.35\textwidth]{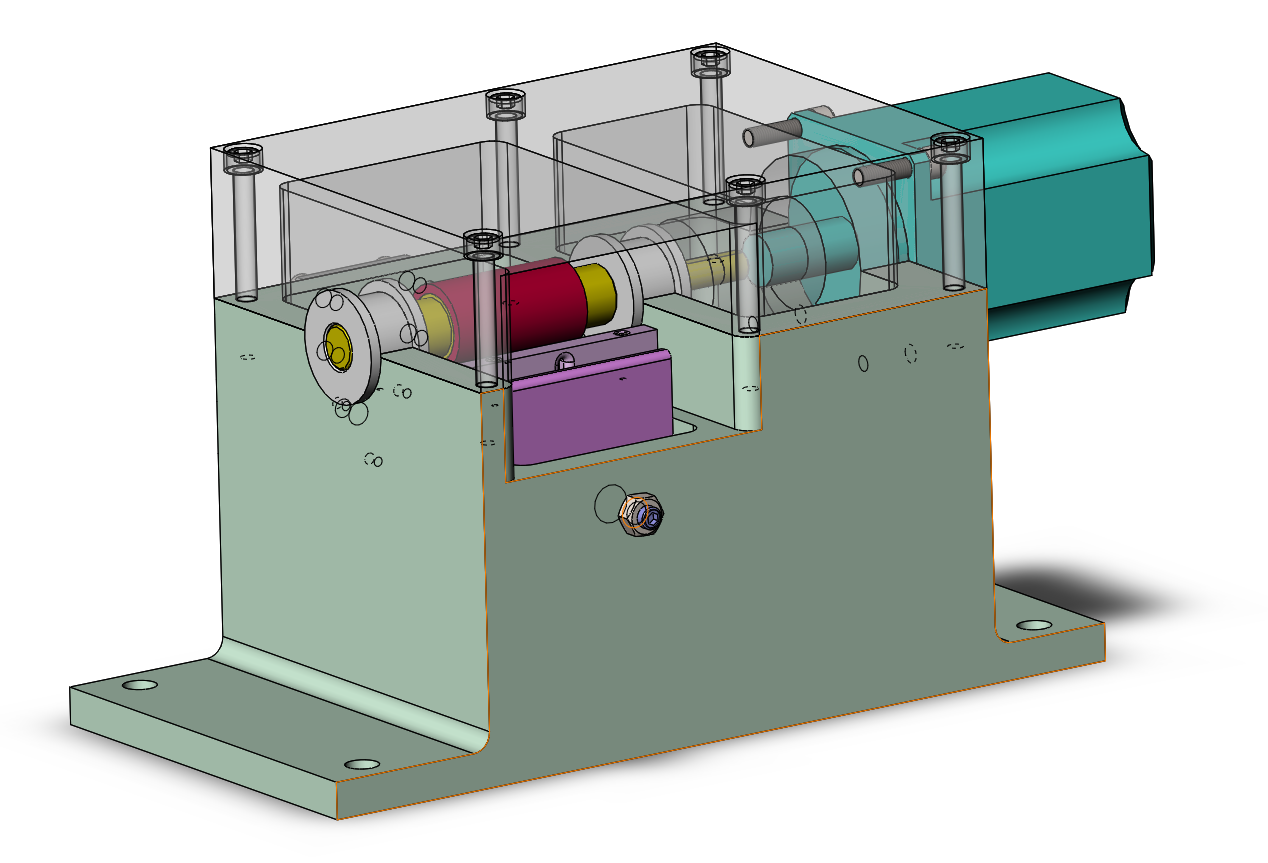}}
    \subfigure[]{    
          \label{fig_TMc}     
          \includegraphics[height=0.35\textwidth]{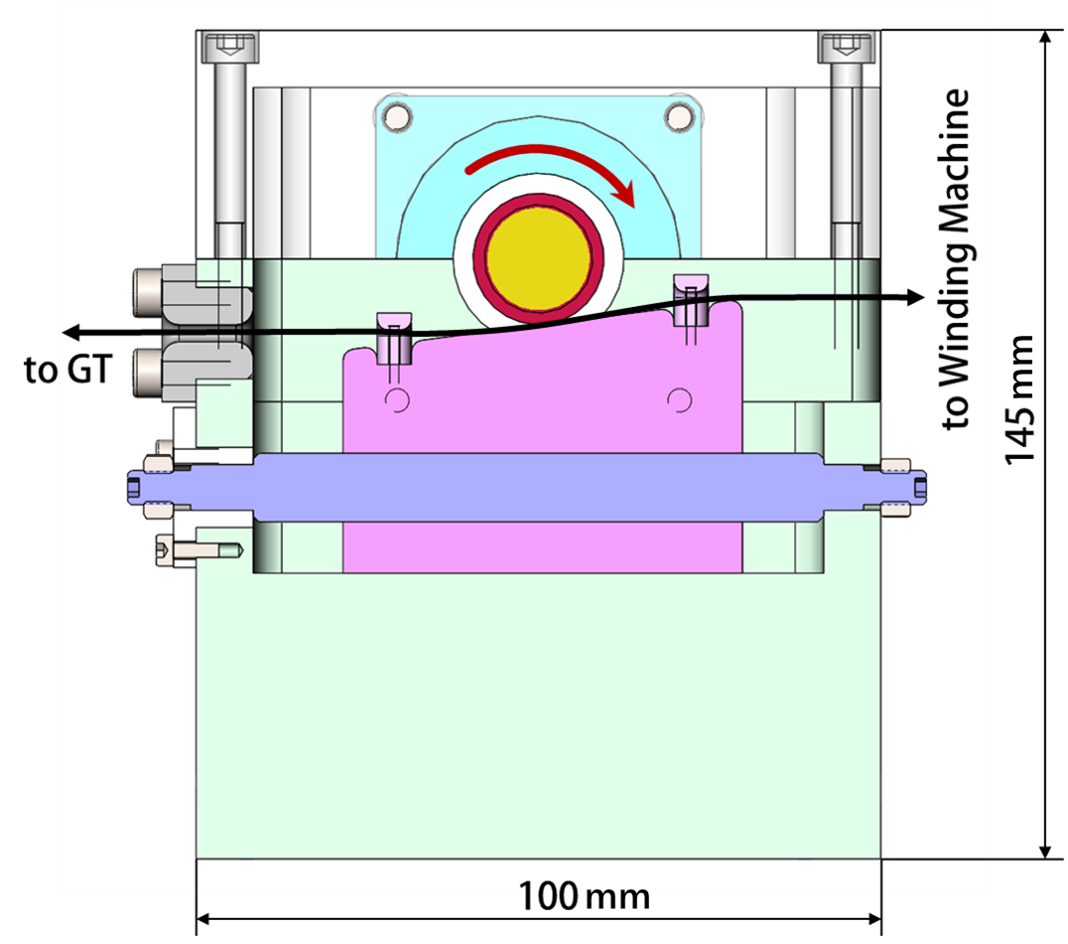}} 
	\caption{Demonstration of the tension maintainer in general view (a) and profile view (b). Yellow: the turning axle. Pink: the holder. Red: damping sleeve.}
	\label{fig:TM}
\end{figure*}

The source carrier is also specially designed as demonstrated in Fig. \ref{fig:source}. The cables are thread into the holes on the shells, wrapped around the cable dowels, and thread out of the shells via the same holes. The dowels are inserted in the inner shell and stainless steel plug. The PTFE plug screwed on inner shell can prevent the source from being lost if the carrier is broken at the stainless steel plug. The assembly of parts from 2 to 7 will be enclosed by the PTFE outer shells in order to reduce the friction. Once the inner and outer shells are fixed by the screws, the retaining key will stop the stainless steel plug from loosening up. This will surely improve the reliability of the source carrier. 

The winding machines are used to manage such a long cable. However, while the cable was releasing during testing, the friction between the cable and cable position limit would prevent the cable from moving smoothly. Thus, the cable in the winding machine would go slack little by little and cause inaccurate positioning as such slack accumulated. An obvious solution was to tighten the cable on the releasing side. However, this seemed contradictory to the requirement of the tension on the cable on the releasing side, considering the friction between tube and cable. The friction between tube and cable will be too big if the tension on the releasing end is not slack (\ref{section:sec-friction}). Therefore, two tension maintainers are applied to solve this problem. 

In addition to the supporting structures, Fig.\ref{fig:TM} shows the tension maintainer
which is composed of a turning axle driven by a motor, a damping sleeve coated on the turning axle and a holder. The cable goes through the maintainer while being squeezed by the damping sleeve and the holder. 
The holder can be moved left and right to adjust the pressure between the cable and the damping sleeve. The friction between the cable and damping sleeve pushes the cable towards the tube when the axle is rolling. The axle rolls in the same direction all the time on both retrieving and releasing side. In this way, the cable on the winding machine will always be tight while keeping a low friction between the cable and tube. This design has been tested on the prototype and the tension on cable on the winding spools was maintained over 50 full positioning cycles.

\section{Studies of simulation bias to energy reconstruction}

In reference \cite{GT_1st}, we have discussed the basic calibration strategy and analysis algorithm of GTCS. 
In order to correct the spatial energy non-uniformity, the CLS, the ACU and the GTCS will work together to map the energy response in CD via 258 designated calibration points. Then, a non-uniformity correction function will be fitted with the measured data \cite{Calib_coll}.
GTCS will provide the boundary condition for the fitting of the non-uniformity correction function. From Fig.\ref{fig_CB}, it can be found that the energy response of GTCS is heavily influenced by the shadowing effect of the chimney and connection bars. In order to save time, less calibration points are preferred. So a model of the boundary condition will be acquired from simulation. Then, the boundary condition will be reconstructed with the model and the energy response on 11 points measured by GTCS.

%is important in the boundary condition reconstruction, because the energy response of GTCS is heavily influenced by the light blocking effect of the chimney and connection bars. With the help of the whole calibration system, JUNO is expected to reach an energy resolution of $3\%/\sqrt{E}$. 

This model is based on the assumption that the simulation should be consistent with the data measured in the final experiment. In another word, it is expected to get the same nPE with the same source at the same position in simulation and in experiment. Otherwise, the reconstructed boundary condition will be influenced by the model. Therefore, a simulation tuning is essential at the beginning of the data taking to remove the simulation bias, which is the difference between the nPE acquired from the simulation and the experiment. However, what if the simulation bias can't be removed because of a failed simulation tuning or systematic error? In this condition, a bias between the energy responses of GTCS in simulation and in experiment data will always exist and thus affect the fitting of the energy response non-uniformity correction, and then the anti-neutrino energy measurement. Therefore, it is necessary to study the influence of the GTCS simulation bias. That is helpful to the simulation tuning and error control.

Since JUNO is still under construction, a feasible way to understand this issue within simulation is to follow the normal energy non-uniformity correction while modifying the boundary condition model manually. The energy resolution and the energy scale uncertainty were used to evaluate the influence. We modified the GTCS model in two ways, one was to introduce a global bias to the energy response. The other one was to use different interpolation models among GTCS calibration points. 
The simulation was still based on the JUNO offline simulation software SNIPER \cite{SNIPER}.

%\subsection{Assurance to biased model}
%\subsection{Dependence on Model's Reconstruction Bias}
\subsection{Impact of an global GTCS bias}

\begin{figure*}[h]
	\centering
    \subfigure[]{    
          \label{fig_Modela}     
          \includegraphics[width=7cm]{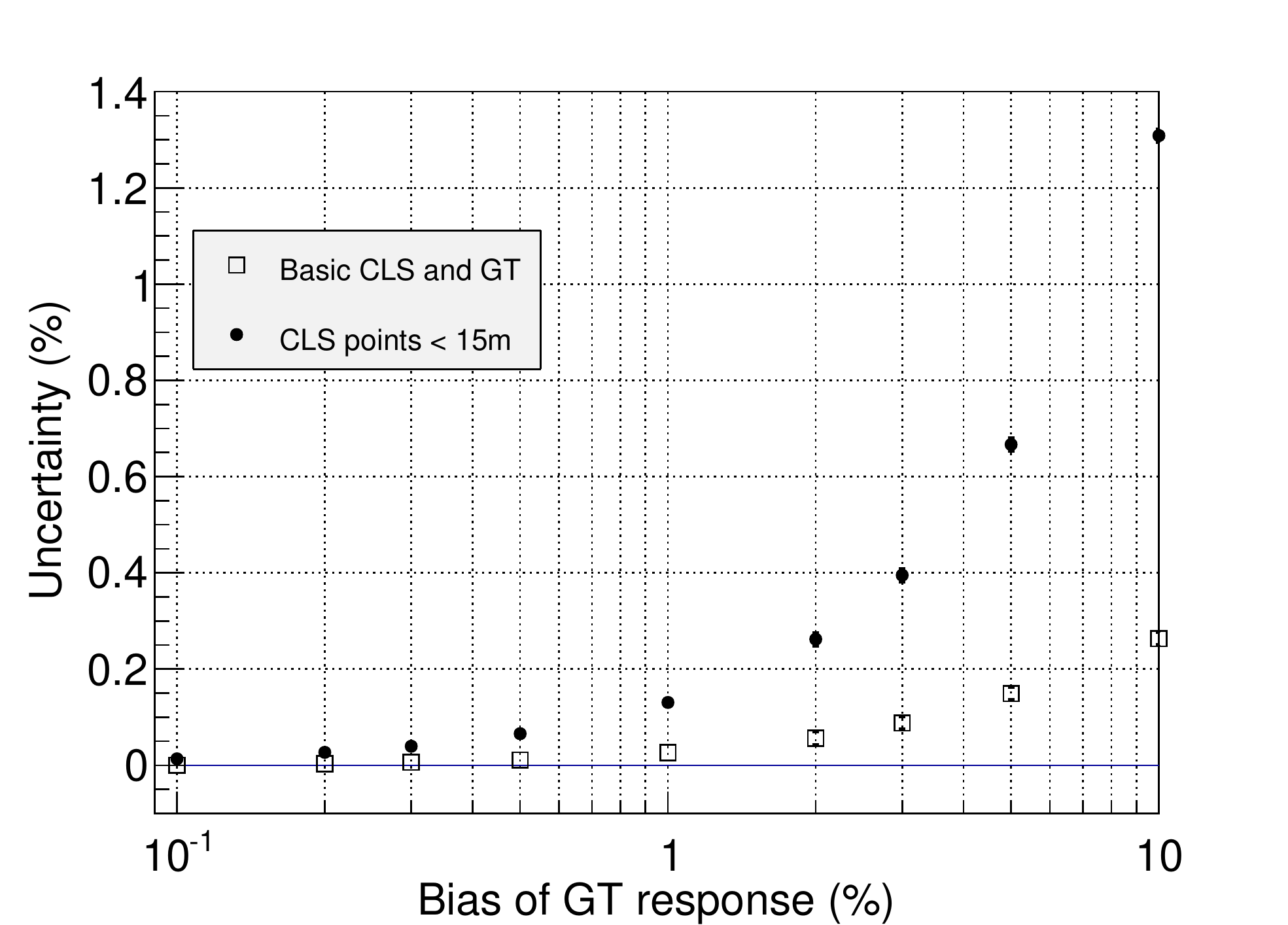}} 
    \subfigure[]{    
          \label{fig_Modelb}     
          \includegraphics[width=7cm]{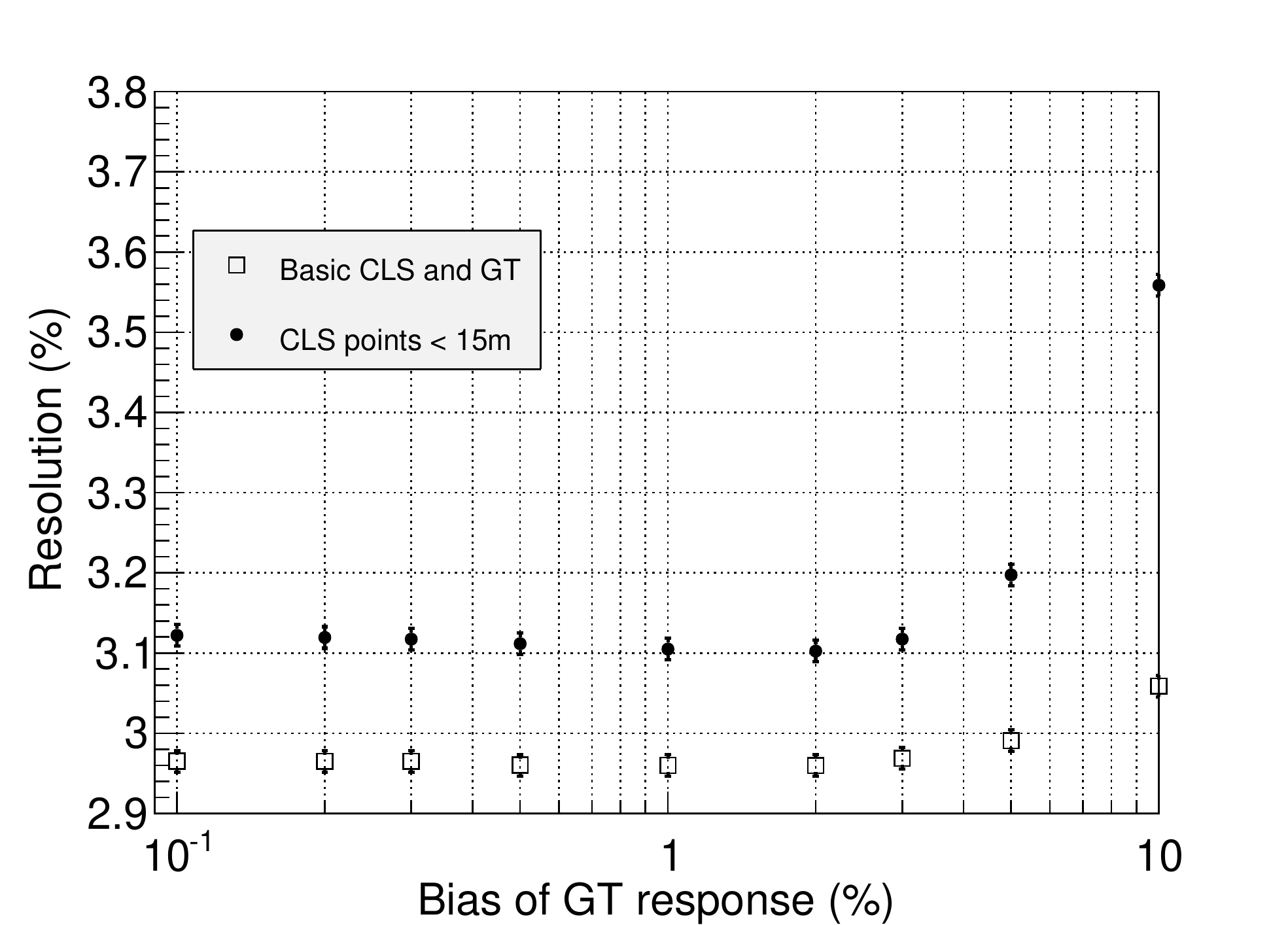}}
	\caption{(a) Impacts on energy scale uncertainty, which is the energy reconstruction mean value bias from the real energy. (b) Impacts on energy  resolution. The open symbols plot the performances with basic design of GTCS and CLS. The solid symbols plot the performances when CLS calibration points are constrained in volume when R < 15 m. The uncertainty is less than 0.1\% and the energy resolution is almost constant,if the model bias is less than 3\%. }
	\label{fig:OverallDrift}
\end{figure*}

The simulation tuning is usually performed at the beginning of data taking to remove the inconsistency between data and simulation. 
Sometimes, the inconsistency of GTCS will be left in order to keep the general consistency of the detector energy response. 
In this situation, the GTCS model from simulation will be globally biased comparing to the real data. 
How much bias is acceptable? To answer this question, we modified the simulated model with a global bias manually and followed the non-uniformity correction procedures. 
The performances of energy reconstruction with differently-biased GTCS models are demonstrated in Fig.\ref{fig:OverallDrift}. The energy scale uncertainty and resolution will increase with the model bias, as expected. 
The open symbols plot the performances of detector with basic CLS and GT design, in which the CLS could deploy source as designed. As the open symbols show, when the bias of GTCS model is smaller than 3\%, the energy resolution does not change much, and the energy scale uncertainty is less than 0.1\%. 
Therefore, considering the optical parameter tuning and systematic error, the simulated boundary condition of GTCS can accept a global bias less than 3\%. 

At the same time, the solid symbols plot the performance when CLS source is constrained in the volume with R < 15 m. Less CLS points could be reached in this condition. 
It can be found that the performance goes bad. This implies the necessity to send the CLS source to the boundary as close as possible. 

\subsection{Dependence on the interpolation functions among GTCS calibration points}

\begin{figure}[h]
\centering
\subfigure[Linear interpolation among the peak and valley points.]{
\includegraphics[width=7cm]{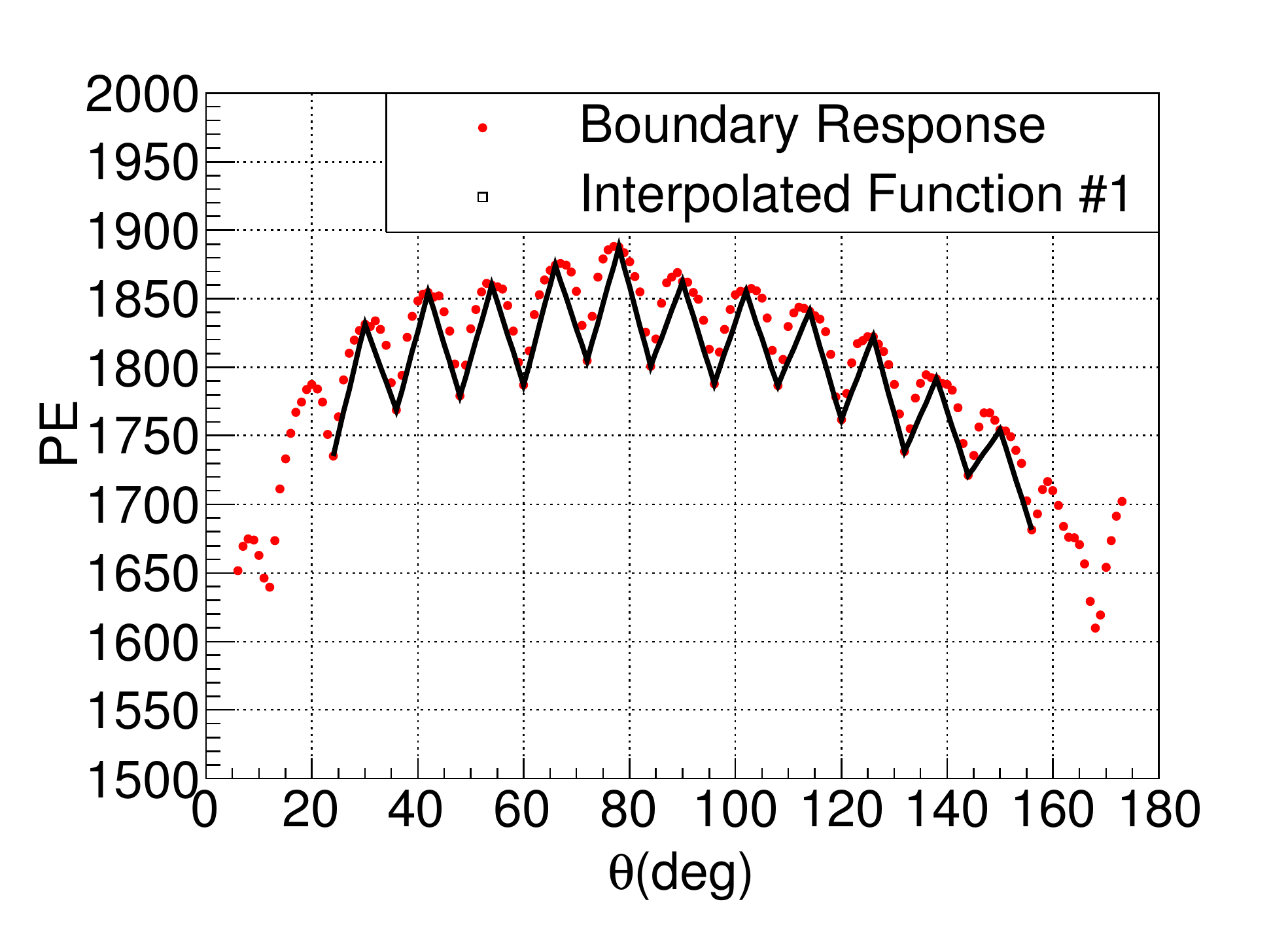}
}
\quad
\subfigure[Linear interpolation among peak points only.]{
\includegraphics[width=7cm]{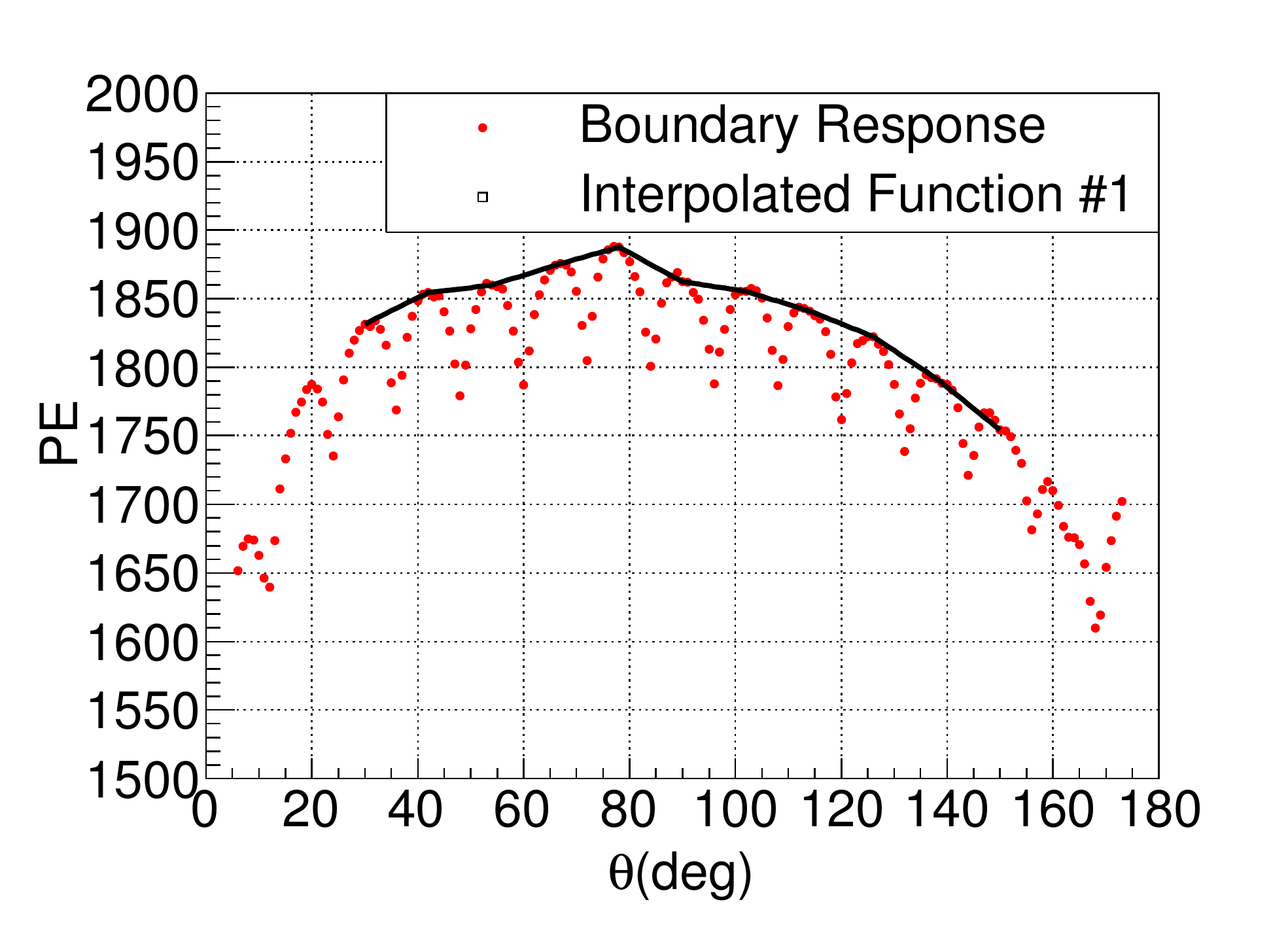}
}
\quad
\subfigure[Constant function using the maximum response.]{
\includegraphics[width=7cm]{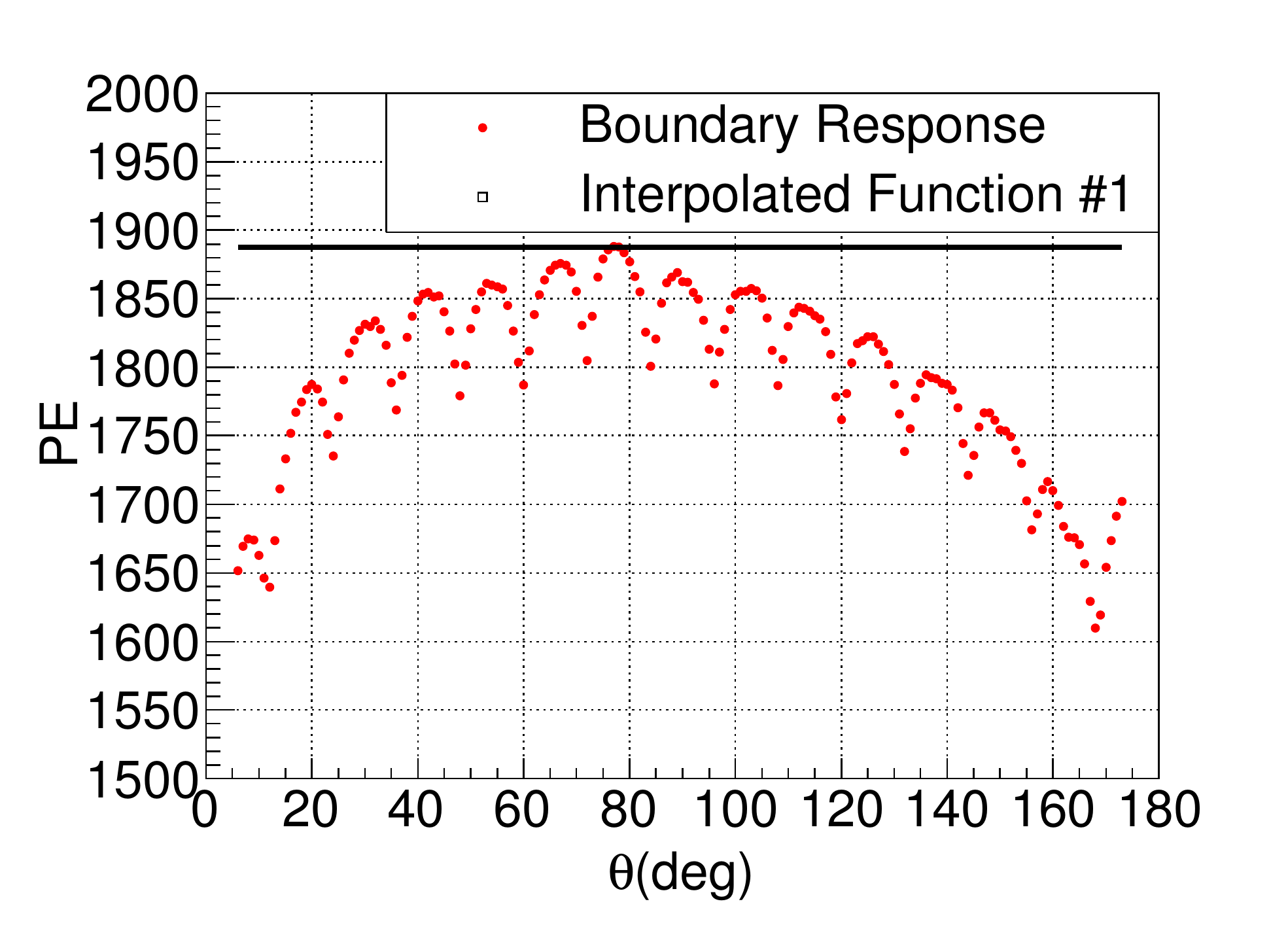}
}
\quad
\subfigure[Constant function using the minimum response.]{
\includegraphics[width=7cm]{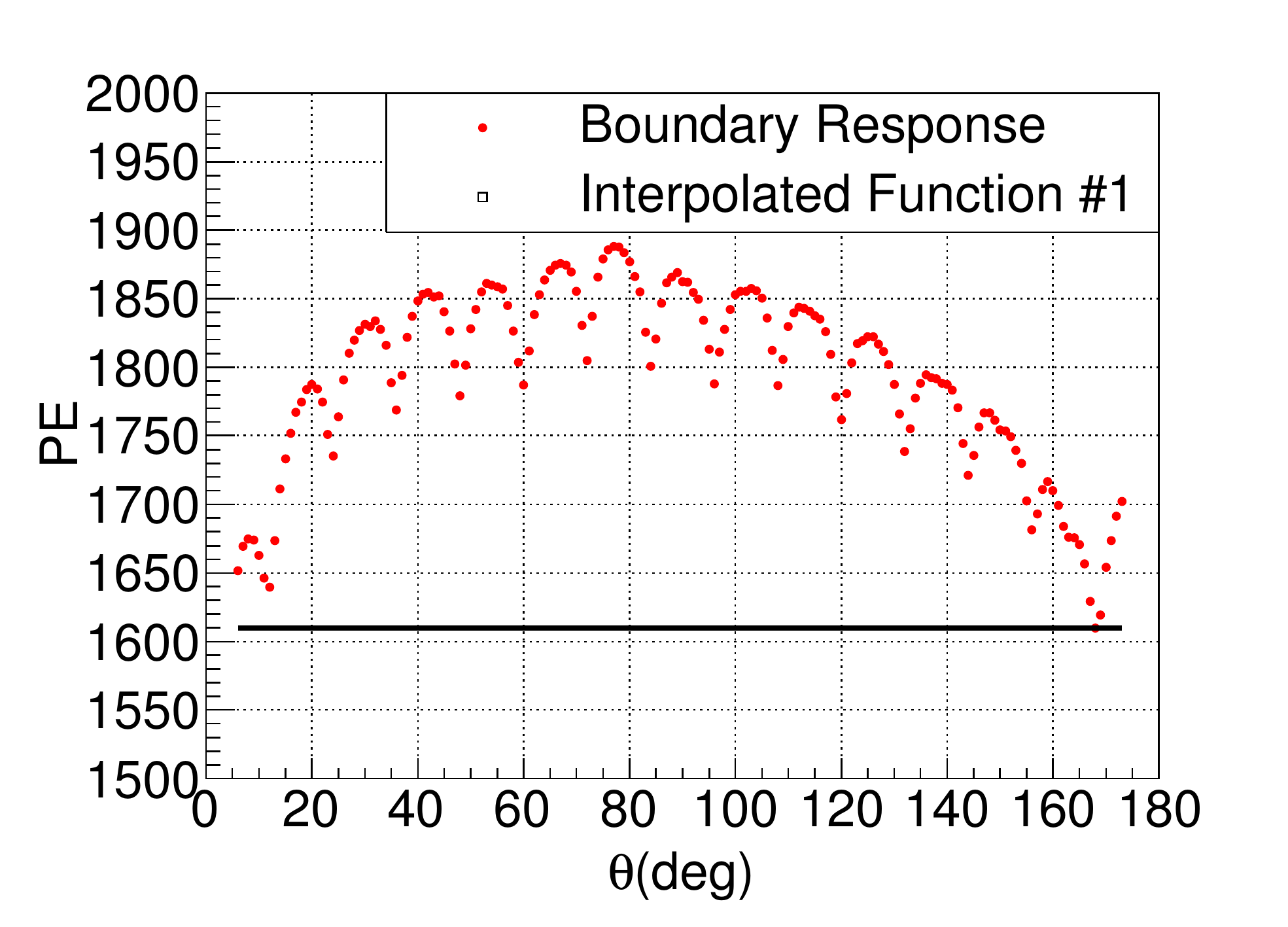}
}
\caption{Four interpolation functions are assumed to be the boundary conditions. The red lines are the true response of $^{40}K$ in GTCS, while the black lines represent the four different interpolation functions.}
\label{fig:IgnoreShape}
\end{figure}

Another potential situation is that the simulation failed to predict the model curve due to unknown reasons. The only reasonable choice for us is to do interpolation with the data and provide an approximate boundary condition for the non-uniformity correction function. 
In Fig.\ref{fig:IgnoreShape}, four approximations were studied: (1) linear interpolation among the peak and valley points; (2) linear interpolation among peaks; (3) a constant function using the maximum response; (4) a constant function using the minimum response. 
Their impacts on the energy reconstruction are summarized in Table \ref{tab:Shape}. The forth method has the worst performance, while all the other three methods can meet the requirements that the energy resolution has to be less than 3\%. 
Obviously, the linear interpolation method is better than the constant model, and should be applied to the non-uniformity correction if needed. 

\begin{table}[]
\centering
\caption{\label{tab:Shape} Performance of different interpolation function}
\begin{tabular}{c|c|c|c}
\Xhline{1.2pt}
  & Interpolation Method                    & Energy Scale Uncertainty & Energy Resolution \\
\Xhline{1.2pt}
1 & Interpolation among peaks and valleys & 0.043\%                  & 2.97\%            \\
\hline
2 & Interpolation among peaks             & 0.094\%                  & 2.98\%            \\
\hline
3 & Constant model using the maximum            & 0.057\%                  & 2.99\%            \\
\hline
4 & Constant model using the minimum            & 0.152\%                  & 3.19\%            \\
\Xhline{1.2pt}
\end{tabular}
\end{table}

\section{Summary}

In this paper, we have introduced the construction of the GTCS. The performance tests were made with the full size system. Finally, the calibration algorithm was also studied when the simulation was inconsistent with the experimental data. Based on these developments, the GTCS is ready to be deployed on the real JUNO detector.

\acknowledgments

This research is supported by the "Strategic Priority Research Program" of the Chinese Academy of Sciences (Grant No. XDA10000000).

% We suggest to always provide author, title and journal data:
% in short all the information that clearly identify a document.

\end{document}